\documentclass[journal,twoside]{IEEEtran}

\usepackage{color}
\usepackage{bbm}
\usepackage{amsmath,amsfonts}
\usepackage{dsfont}
\usepackage{graphicx}
\usepackage[hang]{subfigure}
\usepackage{citesort}
\usepackage{mathtools}
\usepackage{bbm}
\usepackage{fixmath}

\usepackage{ntheorem}
\theoremstyle{plain}
\theoremheaderfont{\itshape}\theorembodyfont{\itshape}
\theoremseparator{.}
\newtheorem{theorem}{Theorem}
\newtheorem{corollary}{Corollary}

\newtheorem{lemma}{Lemma}

\theoremheaderfont{\itshape}\theorembodyfont{\upshape}
\theoremseparator{.}
\newtheorem{definition}{Definition}
\newtheorem{remark}{Remark}

\renewcommand{\vec}[1]{\boldsymbol{#1}}
\newcommand{\E}{\mathbb{E}}
\newcommand{\prob}{\mathbb{P}}

\newcommand{\N}{\mathbb{N}}
\newcommand{\R}{\mathbb{R}}

\newcommand{\sC}{\mathcal{C}}

\newcommand{\sG}{\mathcal{G}}
\newcommand{\sI}{\mathcal{I}}
\newcommand{\sJ}{\mathcal{J}}

\newcommand{\sM}{\mathcal{M}}
\newcommand{\sP}{\mathcal{P}}

\newcommand{\sS}{\mathcal{S}}

\newcommand{\sU}{\mathcal{U}}

\newcommand{\sX}{\mathcal{X}}
\newcommand{\sY}{\mathcal{Y}}
\newcommand{\sZ}{\mathcal{Z}}

\newcommand{\cwc}{\overline{\mathfrak{W}}}
\newcommand{\cc}{\overline{\mathcal{W}}}
\newcommand{\ccv}{\overline{\mathcal{V}}}
\newcommand{\Ccc}{C_S(\cwc)}
\newcommand{\Ccco}{C_S(\cwc_1)}
\newcommand{\Ccct}{C_S(\cwc_2)}
\newcommand{\Cccn}{C_S(\cwc_n)}
\newcommand{\Wso}{W_{s_1}}
\newcommand{\Wst}{W_{s_2}}
\newcommand{\Vso}{V_{s_1}}
\newcommand{\Vst}{V_{s_2}}
\newcommand{\Wtil}{\widetilde{W}}
\newcommand{\Vtil}{\widetilde{V}}
\newcommand{\Ptil}{\widetilde{P}}
\newcommand{\avc}{\mathcal{W}}
\newcommand{\avcv}{\mathcal{V}}
\newcommand{\avwc}{\mathfrak{W}}
\newcommand{\Cdet}{\sC}
\newcommand{\Cran}{\sC_{\text{CR}}}
\newcommand{\Cavc}{C_S(\avwc)}
\newcommand{\Cavco}{C_S(\avwc_1)}
\newcommand{\Cavct}{C_S(\avwc_2)}
\newcommand{\Cavcn}{C_S(\avwc_n)}
\newcommand{\Cavcs}{C_S(\avwc^*)}
\newcommand{\Cavccr}{C_{S,\text{CR}}(\avwc)}
\newcommand{\Cavccrlambda}{C_{S,\text{CR}}(\avwc(\lambda))}

\begin{document}

\title{On the Continuity of the Secrecy Capacity of Compound and Arbitrarily Varying Wiretap Channels}
\author{Holger Boche,~\IEEEmembership{Fellow,~IEEE},
		Rafael F. Schaefer,~\IEEEmembership{Member,~IEEE}, and
		H. Vincent Poor,~\IEEEmembership{Fellow,~IEEE}
	\thanks{Manuscript received September 16, 2014; revised March 22, 2015; accepted July 20, 2015. The work of H.~Boche was supported by the German Ministry of Education and Research under Grant 01BQ1050. The work of R.~F.~Schaefer was supported by the German Research Foundation under Grant WY 151/2-1. The work of H.~V.~Poor was supported by the United States National Science Foundation under Grant CMMI-1435778. This work was presented in part at IEEE-ICC, London, U.K., 2015, and in part at the Industrial Board Meeting and the Review Meeting of the Quantum Repeater Project of the German Ministry of Education and Research (BMBF), Bonn, Germany, May 21-23, 2014. The associate editor coordinating the review of this manuscript and approving it for publication was Prof. Y.-W. Peter Hong.}
	\thanks{H.~Boche is with the Lehrstuhl für Theoretische Informationstechnik, Technische Universität München, Munich 80290, Germany (e-mail: boche@tum.de).}
	\thanks{R.~F.~Schaefer and H.~V.~Poor are with the Department of Electrical Engineering, Princeton University, Princeton, NJ 08544 USA (e-mail: rafaelfs@princeton.edu; poor@princeton.edu).}
	\thanks{Digital Object Identifier 10.1109/TIFS.2015.2465937}
}
\IEEEoverridecommandlockouts
\maketitle

\markboth{IEEE Transactions on Information Forensics and Security}{Boche \MakeLowercase{\textit{et al.}}: On the Continuity of the Secrecy Capacity of Compound and AVWC\MakeLowercase{s}}

\begin{abstract}
The wiretap channel models secure communication between two users in the presence of an eavesdropper who must be kept ignorant of transmitted messages. The performance of such a system is usually characterized by its secrecy capacity which determines the maximum transmission rate of secure communication. In this paper, the issue of whether or not the secrecy capacity is a \emph{continuous} function of the system parameters is examined. In particular, this is done for channel uncertainty modeled via compound channels and arbitrarily varying channels, in which the legitimate users know only that the true channel realization is from a pre-specified uncertainty set. In the former model, this realization remains constant for the entire duration of transmission, while in the latter the realization varies from channel use to channel use in an unknown and arbitrary manner. These models not only capture the case of channel uncertainty, but are also suitable for modeling scenarios in which a malicious adversary jams or otherwise influence the legitimate transmission. The secrecy capacity of the \emph{compound wiretap channel} is shown to be robust in the sense that it is a continuous function of the uncertainty set. Thus, small variations in the uncertainty set lead to small variations in secrecy capacity. On the other hand, the deterministic secrecy capacity of the \emph{arbitrarily varying wiretap channel} is shown to be \emph{discontinuous} in the uncertainty set meaning that small variations can lead to dramatic losses in capacity. 
\end{abstract}


\section{Introduction}
\label{sec:introduction}

In current communication systems, there is usually an architectural separation between error correction and data encryption. The former is typically realized at the physical layer, transforming the noisy communication channel into a reliable ``bit pipe.'' The latter is implemented on top of that by applying cryptographic principles. 

In recent years, \emph{information theoretic approaches to security} have been intensively examined as a complement to such cryptographic techniques. Such approaches establish reliable communication and data confidentiality jointly at the physical layer by taking the properties of the noisy channel into account. This line of study was initiated by Wyner, who introduced the wiretap channel in \cite{Wyner75WiretapChannel}, and subsequently generalized by Csisz\'ar and K\"orner to the broadcast channel with confidential messages \cite{CsiszarKoerner78BroadcastChannelsConfidentialMessages}. Recently, this area of research has drawn considerable attention since it provides a promising approach to achieve security and to embed secure communication into wireless networks; see for example \cite{Liang09InformationTheoreticSecurity,Liu10SecuringWirelessCommunications,Jorswieck10SecrecyPhysicalLayer,Bloch11InformationTheoreticSecrecy,Zhou13PhysicalLayerSecurityBook,SchaeferBoche14SPM} and references therein. Not surprisingly, it therefore has also been identified by operators and national agencies as a key technique to secure future communication systems \cite{Telekom10TechnologyRadar,Helmbrecht08ITSecurityICT,ITU2014TactileInternet}.

These studies are in particular crucial for wireless communication systems, since they are inherently vulnerable to eavesdropping due to the open nature of the wireless medium. Indeed, transmitted signals are received by intended users but are also easily eavesdropped upon by non-legitimate receivers. Many of the previous studies have in common that all channels (including those to the eavesdropper) are assumed to be perfectly known to all users and fixed during the entire duration of transmission. However, in practical systems channel state information (CSI) will always be limited due to the nature of the wireless channel and estimation/feedback inaccuracy. Furthermore, malevolent eavesdroppers will not provide any information about their channels to legitimate users to make eavesdropping even harder. Accordingly, limited CSI (especially for the eavesdropper channel) must be assumed to ensure reliability and data confidentiality. A recent survey on secure communication under channel uncertainty and adversarial attacks can be found in \cite{SchaeferBochePoorXXProcIEEE}.

A first step in the direction of more realistic CSI assumptions is given by the concept of a \emph{compound channel} \cite{Blackwell59Compound,Wolfowitz60SimultaneousChannels}. Here it is assumed that the actual channel realization is unknown. Rather, it is only known to the legitimate users that the true realization belongs to a known set of channels (uncertainty set) and that it remains constant during the entire duration of transmission. Accordingly, secure communication over compound channels is of great importance. The \emph{compound wiretap channel} has been studied in \cite{Liang09CompoundWiretapChannels,Bjelakovic13CompoundWiretap,Ekrem10MIMOCompoundWiretap,Khisti11InterferenceAlignmentMIMOCompoundWiretap,SchaeferLoyka13SecrecyMIMOCompound} and the compound broadcast channel with confidential messages in \cite{Kobayashi09CompoundMIMOBCC,Schaefer14CompoundBCC}. Despite these efforts, a general single-letter characterization of the secrecy capacity remains unknown until now (if it exists at all). Such a description is only known for special cases such as degraded channels \cite{Liang09CompoundWiretapChannels,Bjelakovic13CompoundWiretap} or certain multiple-input multiple-output (MIMO) Gaussian channels \cite{SchaeferLoyka13SecrecyMIMOCompound}. For the general case, only a multi-letter description of the secrecy capacity has been established so far \cite{Bjelakovic13CompoundWiretap}.

The quality of CSI is further weakened by the concept of \emph{arbitrarily varying channels (AVCs)} \cite{Blackwell60AVC,Ahlswede78EliminationCorrelationAVC,CsiszarNarayan88AVCRevisited}. In addition to the assumption that the actual channel realization is known only to be from a known uncertainty set, it is further assumed that this realization may vary from channel use to channel use in an arbitrary and unknown manner (in contrast to compound channels in which it remains constant). The corresponding \emph{arbitrarily varying wiretap channel (AVWC)} has been studied in \cite{MolavianJazi09ArbitraryJamming,Bjelakovic13AVWiretap,BocheSchaefer13Superactivation,BocheSchaeferPoor14AVWCSecrecyMeasures,WieseXXAVWC,NoetzelXXAVWC,SchaeferBochePoorXXSuperactivationUniqueFeature} and it has been shown that it makes a difference in this case whether unassisted or common randomness (CR) assisted codes are used. In particular, the deterministic secrecy capacity may be zero (if the channel possesses the so-called property of symmetrizability as precisely defined later), while the CR-assisted secrecy capacity is non-zero. In \cite{Bjelakovic13AVWiretap,WieseXXAVWC} a complete characterization of the relation between the deterministic and CR-assisted secrecy capacity is established; however, a single-letter characterization of the CR-assisted secrecy capacity itself remains open and only a multi-letter description has been established \cite{WieseXXAVWC}. A Gaussian MIMO wiretap channel where the noiseless eavesdropper channel is arbitrarily varying is considered in \cite{HeYener14MIMOWiretap}. An achievable secrecy rate is derived and the secrecy degrees of freedom are established, while its capacity remains unknown.

These concepts of compound and arbitrarily varying channels not only capture the case of channel uncertainty, but are also suitable for modeling scenarios with active adversaries. For example, such an adversary may be able to maliciously influence the channel conditions by controlling which channel realization governs the transmission. Since the legitimate transmitter and receiver usually have no knowledge about the strategy or the intention of the adversary, they have to choose their encoding-decoding functions in such a way that they work for all possible channel realizations simultaneously. Thus, such attacks can be perfectly modeled by compound channels. It becomes even worse for more powerful adversaries, which may be able to jam the legitimate transmission. Again, having no knowledge about the particular jamming strategy, the legitimate users have to be prepared for a channel that may vary in an unknown and arbitrary manner from channel use to channel use. This is the AVWC, which has been analyzed in \cite{BocheSchaefer13Superactivation} in this context. In particular, the optimal jamming strategy of the adversary has been identified and it is shown that it differs depending on whether the adversary has access to the common randomness or not. An eavesdropper that can control its channel state in a Gaussian two-way wiretap setup is studied in \cite{He11SecrecyEavesdrooperControlsChannel}. Achievable secrecy rates are derived based on cooperative jamming.

The analysis in this paper is driven by the following observation: Obviously, the secrecy capacities of the compound wiretap channel and the AVWC depend on the underlying uncertainty set. Now in general, the performance of a communication system (in our case the secrecy capacity) should depend in a \emph{continuous} way on the system parameters (in particular the uncertainty set). Since, if small changes in the parameters would lead to dramatic losses in performance, the approach at hand will most likely not be used. Indeed, one is interested in approaches that are robust against such variations in the sense that small variations in the uncertainty set result in small variations in the secrecy capacity. Such a continuous dependency is in particular desirable in the context of active adversaries who can influence the system parameters in a malicious way. Surprisingly, the question of continuity of capacities for classical communication scenarios is rarely discussed. However, for the quantum case, there has been some work. Continuity of capacities has been studied in \cite{Leung09ContinuityQuantumCapacities} for quantum channels and in \cite{BocheNotzel14PositivityAVQC} for arbitrarily varying quantum channels. 

In Section \ref{sec:cc} we introduce the compound wiretap channel and a distance concept to measure how ``close'' two compound wiretap channels are. Then in Section \ref{sec:cc_continuity}, we show that the corresponding secrecy capacity is a continuous function of the uncertainty set. Thus, for compound channels, small variations in the uncertainty set result only in small variations of the secrecy capacity. Interpreting the uncertainty set as the strategy space of an adversary, this shows that secure communication over compound channels is robust against changes in the adversary's strategies. 

In Section \ref{sec:avc} we introduce the AVWC and study its secrecy capacity in Section \ref{sec:avc_continuity}. While the secrecy capacity of the compound wiretap channel is continuous in the uncertainty set, we see that the unassisted secrecy capacity of the AVWC can be discontinuous in the uncertainty set. The practical relevance of this observation is that such unassisted schemes might not be robust in the sense that small variations can lead to dramatic losses in secrecy capacity. In particular in the context of active adversaries this means that small changes in the adversary's strategy can lead to completely different behavior of the system. In Section \ref{sec:robust} we analyze the behavior of codes that achieve weak secrecy and show that such codes are robust in the information leakage. This means that a code that realizes weak secrecy for a certain eavesdropper AVC is also good for all such AVCs in a certain neighborhood. Finally, we conclude the paper in Section \ref{sec:conclusion}.

\subsection*{Notation}

Discrete random variables are denoted by capital letters and their realizations and ranges by lower case and script letters, respectively; all logarithms and information quantities are taken to the base 2; $\N$ and $\R_+$ are the sets of non-negative integers and non-negative real numbers; $(0,1)$ and $[0,1]$ are the open and closed intervals between $0$ and $1$; $I(\cdot;\cdot)$ is the mutual information and $H(\cdot)$ and $H_2(\cdot)$ are the traditional entropy and binary entropy functions; the notation $H(\cdot\|P_{XY})$ and $I(\cdot;\cdot\|P_{XY})$ means that the entropy and mutual information are evaluated according to the underlying probability distribution $P_{XY}$; $X-Y-Z$ denotes a Markov chain of random variables $X$, $Y$, and $Z$ in this order; $\prob\{\cdot\}$ is the probability of an event; $\sP(\sX)$ denotes the set of all probability distributions on $\sX$ and $\E_X[\cdot]$ is the expectation with respect to $X$; $\|P_X-Q_X\|$ is the total variation distance between probability distributions $P_X$ and $Q_X$ on $\sX$ defined as $\|P_X-Q_X\|\coloneqq\sum_{x\in\sX}|P_X(x)-Q_X(x)|$; a positive null sequence $\{a_n\}_{n\in\N}$ is a sequence that satisfies $a_n\searrow0$ as $n\rightarrow\infty$; $\text{lhs} \coloneqq \text{rhs}$ means the value of the right hand side
(rhs) is assigned to the left hand side (lhs), $\text{lhs} \eqqcolon \text{rhs}$ is defined accordingly.

\section{Compound Wiretap Channels}
\label{sec:cc}

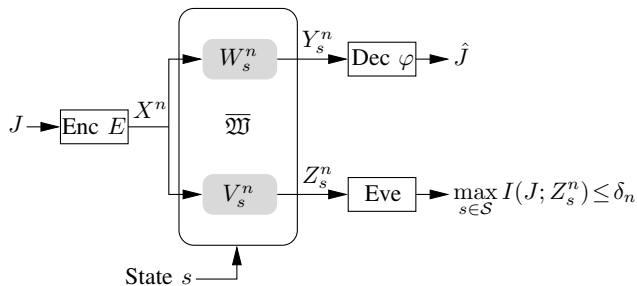
\begin{figure}
  \centering
  	\scalebox{0.9}{\hspace{-2.5cm}
  	\begin{picture}(0,0)%
  	\includegraphics{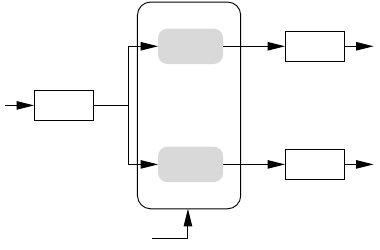}%
  	\end{picture}%
  	\setlength{\unitlength}{3947sp}%
  	\begingroup\makeatletter\ifx\SetFigFont\undefined%
  	\gdef\SetFigFont#1#2#3#4#5{%
  	  \fontfamily{#3}\fontseries{#4}\fontshape{#5}%
  	  \selectfont}%
  	\fi\endgroup%
  	\begin{picture}(3031,1996)(2800,-2208)
  	\put(5816,-1554){\makebox(0,0)[lb]{\smash{{\SetFigFont{8}{9.6}{\rmdefault}{\mddefault}{\updefault}{\color[rgb]{0,0,0}$\displaystyle\max_{s\in\sS}I(J;Z^n_s)\!\leq\!\delta_n$}%
  	}}}}
  	\put(5816,-618){\makebox(0,0)[lb]{\smash{{\SetFigFont{8}{9.6}{\rmdefault}{\mddefault}{\updefault}{\color[rgb]{0,0,0}$\hat{J}$}%
  	}}}}
  	\put(3993,-2151){\makebox(0,0)[rb]{\smash{{\SetFigFont{8}{9.6}{\rmdefault}{\mddefault}{\updefault}{\color[rgb]{0,0,0}$\text{State } s$}%
  	}}}}
  	\put(2815,-1087){\makebox(0,0)[rb]{\smash{{\SetFigFont{8}{9.6}{\rmdefault}{\mddefault}{\updefault}{\color[rgb]{0,0,0}$J$}%
  	}}}}
  	\put(3700,-988){\makebox(0,0)[b]{\smash{{\SetFigFont{8}{9.6}{\rmdefault}{\mddefault}{\updefault}{\color[rgb]{0,0,0}$X^n$}%
  	}}}}
  	\put(4868,-495){\makebox(0,0)[b]{\smash{{\SetFigFont{8}{9.6}{\rmdefault}{\mddefault}{\updefault}{\color[rgb]{0,0,0}$Y^n_s$}%
  	}}}}
  	\put(4868,-1440){\makebox(0,0)[b]{\smash{{\SetFigFont{8}{9.6}{\rmdefault}{\mddefault}{\updefault}{\color[rgb]{0,0,0}$Z^n_s$}%
  	}}}}
  	\put(5324,-616){\makebox(0,0)[b]{\smash{{\SetFigFont{8}{9.6}{\rmdefault}{\mddefault}{\updefault}{\color[rgb]{0,0,0}$\text{Dec } \varphi$}%
  	}}}}
  	\put(5324,-1567){\makebox(0,0)[b]{\smash{{\SetFigFont{8}{9.6}{\rmdefault}{\mddefault}{\updefault}{\color[rgb]{0,0,0}$\text{Eve}$}%
  	}}}}
  	\put(3308,-1100){\makebox(0,0)[b]{\smash{{\SetFigFont{8}{9.6}{\rmdefault}{\mddefault}{\updefault}{\color[rgb]{0,0,0}$\text{Enc } E$}%
  	}}}}
  	\put(4312,-1104){\makebox(0,0)[b]{\smash{{\SetFigFont{8}{9.6}{\rmdefault}{\mddefault}{\updefault}{\color[rgb]{0,0,0}$\cwc$}%
  	}}}}
  	\put(4316,-1572){\makebox(0,0)[b]{\smash{{\SetFigFont{8}{9.6}{\rmdefault}{\mddefault}{\updefault}{\color[rgb]{0,0,0}$V_s^n$}%
  	}}}}
  	\put(4316,-618){\makebox(0,0)[b]{\smash{{\SetFigFont{8}{9.6}{\rmdefault}{\mddefault}{\updefault}{\color[rgb]{0,0,0}$W_s^n$}%
  	}}}}
  	\end{picture}%
  	}
  \caption{Compound wiretap channel. The transmitter encodes the message $J$ into the codeword $X^n=E(J)$ and transmits it over the compound wiretap channel to the legitimate receiver, which has to decode its intended message $\hat{J}=\varphi(Y_s^n)$ for any channel realization $s\in\sS$. At the same time, the eavesdropper has to be kept ignorant of $J$ in the sense that $\max_{s\in\sS}I(J;Z_s^n)\leq\delta_n$.}
  \label{fig:cc}
\end{figure}

We begin with the \emph{compound wiretap channel} in which the actual channel realization is unknown to the legitimate users. It is known only that it is constant during the entire duration of transmission and lies in a known uncertainty set. Furthermore, no prior distribution on the uncertainty set is assumed. In particular, this models scenarios in which an adversary influences the channel conditions by choosing the actual realization unknown to the legitimate users. The compound wiretap channel is depicted in Fig. \ref{fig:cc}.

The main result will be that the secrecy capacity of the compound wiretap channel is continuous in the uncertainty set. The consequence is that the presented approach for secure communication over compound channels is robust against such classes of attacks.

Let $\sX$, $\sY$, and $\sZ$ be finite input and output sets and $\sS$ be an arbitrary state set. Then for given state $s\in\sS$ and input and output sequences $x^n\in\sX^n$, $y^n\in\sY^n$, and $z^n\in\sZ^n$ of length $n$, the discrete memoryless channels to the legitimate receiver and the eavesdropper are given by $W_s^n(y^n|x^n)\coloneqq\prod_{i=1}^nW_s(y_i|x_i)$ and $V_s^n(z^n|x^n)\coloneqq\prod_{i=1}^nV_s(z_i|x_i)$, respectively. 

Then the (marginal) compound channel to the legitimate receiver is defined by the family of channels for all $s\in\sS$ as
\begin{equation*}
	\cc \coloneqq \big\{W_s\big\}_{s\in\sS}.
\end{equation*}
Similarly, we define the compound channel to the eavesdropper as $\ccv \coloneqq \{V_s\}_{s\in\sS}$.

\begin{definition}
\label{def:cc_cc}
The discrete memoryless \emph{compound wiretap channel} is given by the families of pairs of compound channels with common input as
\begin{equation*}
	\cwc \coloneqq \big\{\cc,\ccv\big\} = \big\{W_s,V_s\big\}_{W_s\in\cc,V_s\in\ccv}.
\end{equation*}
\end{definition}

\begin{remark}
\label{rem:uncertaintyset}
Throughout the paper we will also refer to $\cwc$ as the uncertainty set of the compound wiretap channel. Later in Section \ref{sec:cc_distance} we will clarify why it is reasonable to define the uncertainty by the channel matrices $(\{W_s\}_{s\in\sS},\{V_s\}_{s\in\sS})$ and not by the state set $\sS$ itself.
\end{remark}

\subsection{Codes for Compound Wiretap Channels}
\label{sec:cc_codes}

We consider a block code of arbitrary but fixed length $n$. Let $\sJ_n\coloneqq\{1,...,J_n\}$ be the set of confidential messages.

\begin{definition}
\label{def:cc_code}
An $(n,J_n)$-\emph{code} $\Cdet$ consists of a stochastic encoder
\begin{equation}
	E : \sJ_n \rightarrow\sP(\sX^n)
	\label{eq:cc_encoder}
\end{equation}
and a deterministic decoder at the legitimate receiver
\begin{equation}
	\varphi : \sY^n \rightarrow\sJ_n.
	\label{eq:cc_decoder}
\end{equation}
\end{definition}

The encoder $E$ in \eqref{eq:cc_encoder} is allowed to be stochastic. This means that it is specified by conditional probabilities $E(x^n|j)$ with $\sum_{x^n\in\sX^n}E(x^n|j)=1$ for each $j\in\sJ_n$, where $E(x^n|j)$ denotes the probability that the transmitter encodes the message $j\in\sJ_n$ as $x^n\in\sX^n$.

\begin{remark}
\label{rem:cc_code}
For the compound wiretap channel it suffices to consider codes as defined in Definition \ref{def:cc_code}. However, we will see that for the AVWC in Section \ref{sec:avc} we need more sophisticated code concepts based on common randomness; so-called CR-assisted codes, cf. Definition \ref{def:avc_rancode}. In this context, we will then refer to codes of Definition \ref{def:cc_code} as \emph{unassisted codes}.
\end{remark}

\begin{remark}
\label{rem:cc_universal}
Since the true channel realization is unknown to the transmitter and receiver, the encoder \eqref{eq:cc_encoder} and decoder \eqref{eq:cc_decoder} must be universal in the sense that they do not depend on the particular state $s\in\sS$.
\end{remark}

When the transmitter has sent the message $j\in\sJ_n$ and the legitimate receiver has received $y^n\in\sY^n$, its decoder is in error if $\varphi(y^n)\neq j$. Then for an $(n,J_n)$-code $\Cdet$, the average probability of error for channel realization $s\in\sS$ is given by
\begin{align*}
	\bar{e}_n(s\|\Cdet) \coloneqq \frac{1}{|\sJ_n|}\sum_{j\in\sJ_n}\sum_{x^n\in\sX^n}\sum_{y^n:\varphi(y^n)\neq j}W_s^n(y^n|x^n)E(x^n|j).
\end{align*}

To ensure the confidentiality of the message for all channel realizations $s\in\sS$, we require $\max_{s\in\sS}I(J;Z_s^n\|\Cdet)\leq\delta_n$ for some (small) $\delta_n>0$ with $J$ the random variable uniformly distributed over the set of messages $\sJ_n$ and $Z_s^n=(Z_{s,1},Z_{s,2},...,Z_{s,n})$ the output at the eavesdropper for channel realization $s\in\sS$. This criterion is known as \emph{strong secrecy} \cite{Csiszar96SecrecyCapacity,Maurer00WeakToStrongSecrecy} and the motivation behind this is to control the total amount of information leaked to the eavesdropper. This yields the following definition.

\begin{remark}
\label{rem:cc_condition}
Conditioning on the code $\Cdet$ in $I(J;Z_s^n\|\Cdet)$ indicates that the mutual information term is evaluated under this particular kind of code, i.e., the underlying joint probability distribution is given by
	$P_{JX^nZ_s^n}(j,x^n,z^n) = V_s^n(z^n|x^n)E(x^n|j)\frac{1}{|\sJ_n|}$
since the messages are assumed to be uniformly distributed. Note that for the compound wiretap channel, this notation is dispensable since we only deal with one class of codes, cf. Definition \ref{def:cc_code}. The notation becomes crucial for AVWCs, where different code concepts are used.
\end{remark}

\begin{definition}
\label{def:cc_achievable}
A rate $R>0$ is said to be an \emph{achievable secrecy rate} for the compound wiretap channel if for any $\tau>0$ there exist an $n(\tau)\in\N$, positive null sequences $\{\lambda_n\}_{n\in\N}$, $\{\delta_n\}_{n\in\N}$, and a sequence of $(n,J_n)$-codes $\{\Cdet_n\}_{n\in\N}$ such that for all $n\geq n(\tau)$ we have $\frac{1}{n}\log J_n\geq R - \tau$, $\sup_{s\in\sS}\bar{e}_n(s\|\Cdet_n) \leq \lambda_n$, and $\sup_{s\in\sS}I(J;Z_s^n\|\Cdet_n) \leq \delta_n$. The \emph{secrecy capacity} $\Ccc$ of the compound wiretap channel with uncertainty set $\cwc$ is given by the maximum of all achievable secrecy rates~$R$.
\end{definition}

\subsection{Capacity Results}
\label{sec:cc_capacity}

The compound wiretap channel has been studied in several contexts. In \cite[Theorem 1]{Liang09CompoundWiretapChannels} an achievable secrecy rate for finite uncertainty sets and the weak secrecy criterion is established. The result has been strengthened in \cite{Bjelakovic13CompoundWiretap} and \cite{SchaeferLoyka13SecrecyMIMOCompound} to hold also for strong secrecy and arbitrary (not necessarily finite or countable) uncertainty sets. A natural upper bound on the secrecy capacity is given by the worst-case approach, since the secrecy capacity of the compound wiretap channel cannot exceed the secrecy capacity of the worst wiretap channel in this set \cite[Theorem 2]{Liang09CompoundWiretapChannels}. For degraded channels (where each realization of the eavesdropper channel is a degraded version of each realization of the legitimate channel) it has been shown in \cite[Theorem 3]{Liang09CompoundWiretapChannels} that the secrecy rate in \cite[Theorem 1]{Liang09CompoundWiretapChannels} is actually the secrecy capacity. 

Although a single-letter expression for the secrecy capacity that holds in the general, non-degraded, case is still unknown, a multi-letter description was established in \cite[Remark 2]{Bjelakovic13CompoundWiretap}.

\begin{theorem}[\cite{Bjelakovic13CompoundWiretap}]
\label{the:cc_secrecy}
The \emph{secrecy capacity} $\Ccc$ of the compound wiretap channel with uncertainty set $\cwc$ is 
\begin{align}
	&\Ccc =\lim_{n\rightarrow\infty}\frac{1}{n}\max_{U-X^n-(Y_s^n,Z_s^n)} \nonumber \\
	&\qquad\qquad\qquad\times\big(\inf_{s\in\sS}I(U;Y_s^n) - \sup_{s\in\sS}I(U;Z_s^n)\big)
	\label{eq:cc_capacity}
\end{align}
for random variables $U-X^n-(Y_s^n,Z_s^n)$ forming a Markov chain. 
\end{theorem}

\begin{remark}
\label{rem:cc_existence}
Note that the limit in \eqref{eq:cc_capacity} exists and is well defined, cf. \cite[Lemma 5]{Bjelakovic13CompoundWiretap}.
\end{remark}

In the next section we want to use the multi-letter expression \eqref{eq:cc_capacity} in Theorem \ref{the:cc_secrecy} to analyze the dependence of the secrecy capacity on the uncertainty set.

\section{Continuity of Compound Secrecy Capacity}
\label{sec:cc_continuity}

In this section, we analyze the secrecy capacity $\Ccc$ of the compound wiretap channel 
and show that it is a \emph{continuous} function of the uncertainty set $\cwc$. For this purpose, we need a concept to measure the distance between two compound wiretap channels as introduced in the following.

\subsection{Distance between Compound Wiretap Channels}
\label{sec:cc_distance}

Let $(W,V)$ and $(\Wtil,\Vtil)$ be two wiretap channels (with finite input and output alphabets $\sX$, $\sY$, and $\sZ$). We define the distance between two (marginal) channels based on the total variation distance\footnote{Note that it is not important which particular norm is used to define the distance. This follows from the fact the output alphabet $\sY$ is finite so that all norms are equivalent. Choosing a norm other than the total variation distance in \eqref{eq:cc_distance} would only lead to slightly different constants in the results below (e.g. in \eqref{eq:cc_continuity} of Theorem \ref{the:cc_continuity}.).} as 
\begin{subequations}
\label{eq:cc_distance}
\begin{align}
	d(W,\Wtil) &\coloneqq \max_{x\in\sX}\sum_{y\in\sY}\big|W(y|x)-\Wtil(y|x)\big| \\
	d(V,\Vtil) &\coloneqq \max_{x\in\sX}\sum_{z\in\sZ}\big|V(z|x)-\Vtil(z|x)\big|
\end{align}
\end{subequations}
and between the corresponding wiretap channels as
\begin{equation*}
	d\big((W,V),(\Wtil,\Vtil)\big) \coloneqq \max\big\{d(W,\Wtil),d(V,\Vtil)\big\}.
\end{equation*}

Next, we extend this concept to the compound case. Accordingly, let $\cwc_1=(\cc_1,\ccv_1)$ and $\cwc_2=(\cc_2,\ccv_2)$ with index sets $\sS_1$ and $\sS_2$ be two uncertainty sets for compound wiretap channels with marginal compound channels $\cc_i=\{W_{s_i}\}_{s_i\in\sS_i}$ and $\overline{\mathcal{V}}_i=\{V_{s_i}\}_{s_i\in\sS_i}$, $i=1,2$. We define distances between the legitimate compound channels as
\begin{align*}
	d_{1}(\cc_1,\cc_2) &= \sup_{s_2\in\sS_2}\inf_{s_1\in\sS_1}d(\Wso,\Wst) \\
	d_{2}(\cc_1,\cc_2) &= \sup_{s_1\in\sS_1}\inf_{s_2\in\sS_2}d(\Wso,\Wst)
\end{align*}
and between the eavesdropper compound channels as
\begin{align*}
	d_{1}(\ccv_1,\ccv_2) &= \sup_{s_2\in\sS_2}\inf_{s_1\in\sS_1}d(\Vso,\Vst) \\
	d_{2}(\ccv_1,\ccv_2) &= \sup_{s_1\in\sS_1}\inf_{s_2\in\sS_2}d(\Vso,\Vst).
\end{align*}

\begin{definition}
\label{def:cc_distance}
The distance $D(\cwc_1,\cwc_2)$ between two compound wiretap channels with uncertainty sets $\cwc_1$ and $\cwc_2$ is defined as
\begin{equation}
\begin{split}
	D(\cwc_1,\cwc_2) = \max\big\{&d_{1}(\cc_1,\cc_2),d_{2}(\cc_1,\cc_2),\\
	&d_{1}(\ccv_1,\ccv_2),d_{2}(\ccv_1,\ccv_2)\big\}.
	\label{eq:cc_defdistance}
\end{split}
\end{equation}
\end{definition}

Roughly speaking, the distance $D(\cwc_1,\cwc_2)$ between two compound wiretap channels is given by the largest distance in \eqref{eq:cc_distance} for all possible channel realizations in the corresponding uncertainty sets $\cwc_1$ and $\cwc_2$. It characterizes how ``close'' or similar these two compound wiretap channels are. Accordingly, it can also be interpreted as a measure of how well one compound wiretap channel can be approximated by another one.

Further, this definition has the following implications. Let $\cwc$ be an uncertainty set with state set $\sS$ and further let $\{\cwc_n\}_{n\in\N}$ be a sequence with respective state sets $\{\sS_n\}_{n\in\N}$ that satisfy
\begin{equation*}
	\lim_{n\rightarrow\infty}D(\cwc,\cwc_n)=0.
\end{equation*}
Then there exists for every channel realization $(W_{\hat{s}},V_{\hat{s}})$ from the uncertainty set $\cwc=\{\cc,\ccv\}$ with state set $\sS$ a sequence $\{\hat {s}_n\}_{n\in\N}$ with $\hat{s}_n\in\hat{\sS}_n$ and
\begin{align*}
	\lim_{n\rightarrow\infty} \max_{x\in\sX}\sum_{y\in\sY}\big|W_{\hat{s}}(y|x)-W_{\hat{s}_n}(y|x)\big| &=0\\
	\lim_{n\rightarrow\infty} \max_{x\in\sX}\sum_{z\in\sZ}\big|V_{\hat{s}}(z|x)-V_{\hat{s}_n}(z|x)\big| &=0.
\end{align*}
Moreover, for every $\epsilon>0$ there exists an $n_0=n_0(\epsilon)$ such that for all $n\geq n_0$ and all $(W_{\hat{s}},V_{\hat{s}})\in\cwc_n$ there exists a channel realization $(W_{\tilde{s}},V_{\tilde{s}})\in\cwc$ such that
\begin{align*}
	\max_{x\in\sX}\sum_{y\in\sY}\big|W_{\hat{s}}(y|x)-W_{\tilde{s}}(y|x)\big| &<\epsilon\\
	\max_{x\in\sX}\sum_{z\in\sZ}\big|V_{\hat{s}}(z|x)-V_{\tilde{s}}(z|x)\big| &<\epsilon.
\end{align*}

\begin{figure}
  \centering
   	\scalebox{0.9}{\hspace{0.2cm} 
   	\begin{picture}(0,0)%
   	\includegraphics{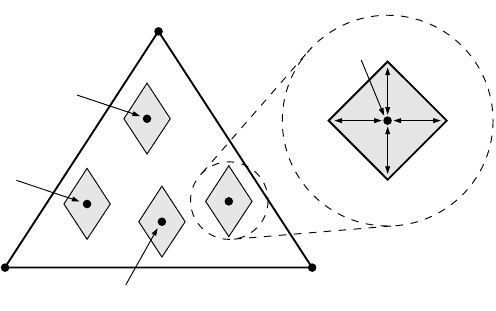}%
   	\end{picture}%
   	\setlength{\unitlength}{3947sp}%
   	\begingroup\makeatletter\ifx\SetFigFont\undefined%
   	\gdef\SetFigFont#1#2#3#4#5{%
   	  \fontfamily{#3}\fontseries{#4}\fontshape{#5}%
   	  \selectfont}%
   	\fi\endgroup%
   	\begin{picture}(3953,2473)(3457,-2340)
   	\put(6403,-263){\makebox(0,0)[b]{\smash{{\SetFigFont{8}{9.6}{\rmdefault}{\mddefault}{\updefault}{\color[rgb]{0,0,0}$W_{s_2}(\cdot|x_1)$}%
   	}}}}
   	\put(3591,-1218){\makebox(0,0)[b]{\smash{{\SetFigFont{8}{9.6}{\rmdefault}{\mddefault}{\updefault}{\color[rgb]{0,0,0}$W_{s_1}(\cdot|x_1)$}%
   	}}}}
   	\put(4073,-548){\makebox(0,0)[b]{\smash{{\SetFigFont{8}{9.6}{\rmdefault}{\mddefault}{\updefault}{\color[rgb]{0,0,0}$W_{s_1}(\cdot|x_2)$}%
   	}}}}
   	\put(4464,-2283){\makebox(0,0)[b]{\smash{{\SetFigFont{8}{9.6}{\rmdefault}{\mddefault}{\updefault}{\color[rgb]{0,0,0}$W_{s_2}(\cdot|x_2)$}%
   	}}}}
   	\put(4736,  6){\makebox(0,0)[b]{\smash{{\SetFigFont{8}{9.6}{\rmdefault}{\mddefault}{\updefault}{\color[rgb]{0,0,0}$(0,0,1)$}%
   	}}}}
   	\put(3497,-2190){\makebox(0,0)[b]{\smash{{\SetFigFont{8}{9.6}{\rmdefault}{\mddefault}{\updefault}{\color[rgb]{0,0,0}$(1,0,0)$}%
   	}}}}
   	\put(5954,-2190){\makebox(0,0)[b]{\smash{{\SetFigFont{8}{9.6}{\rmdefault}{\mddefault}{\updefault}{\color[rgb]{0,0,0}$(0,1,0)$}%
   	}}}}
   	\put(6588,-1091){\makebox(0,0)[lb]{\smash{{\SetFigFont{8}{9.6}{\rmdefault}{\mddefault}{\updefault}{\color[rgb]{0,0,0}$\epsilon$}%
   	}}}}
   	\put(6590,-655){\makebox(0,0)[lb]{\smash{{\SetFigFont{8}{9.6}{\rmdefault}{\mddefault}{\updefault}{\color[rgb]{0,0,0}$\epsilon$}%
   	}}}}
   	\put(6796,-939){\makebox(0,0)[b]{\smash{{\SetFigFont{8}{9.6}{\rmdefault}{\mddefault}{\updefault}{\color[rgb]{0,0,0}$\epsilon$}%
   	}}}}
   	\put(6350,-939){\makebox(0,0)[b]{\smash{{\SetFigFont{8}{9.6}{\rmdefault}{\mddefault}{\updefault}{\color[rgb]{0,0,0}$\epsilon$}%
   	}}}}
   	\put(6808,-1196){\makebox(0,0)[lb]{\smash{{\SetFigFont{8}{9.6}{\rmdefault}{\mddefault}{\updefault}{\color[rgb]{0,0,0}$\sM_{12}$}%
   	}}}}
   	\put(4242,-1792){\makebox(0,0)[lb]{\smash{{\SetFigFont{8}{9.6}{\rmdefault}{\mddefault}{\updefault}{\color[rgb]{0,0,0}$\sM_{11}$}%
   	}}}}
   	\put(4840,-1937){\makebox(0,0)[lb]{\smash{{\SetFigFont{8}{9.6}{\rmdefault}{\mddefault}{\updefault}{\color[rgb]{0,0,0}$\sM_{22}$}%
   	}}}}
   	\put(4738,-1089){\makebox(0,0)[lb]{\smash{{\SetFigFont{8}{9.6}{\rmdefault}{\mddefault}{\updefault}{\color[rgb]{0,0,0}$\sM_{21}$}%
   	}}}}
   	\end{picture}%
   	}
  \caption{Output probability distributions on $\sY$. The corner points correspond to the stochastic matrices with $\prob\{Y=y_i\}=1$, $i=1,2,3$. The shaded areas are $\epsilon$-regions where all stochastic matrices are close to a specific stochastic matrix according to \eqref{eq:cc_distance}. Note that the shape of these regions depend on the applied norm.}
  \label{fig:compound}
\end{figure}
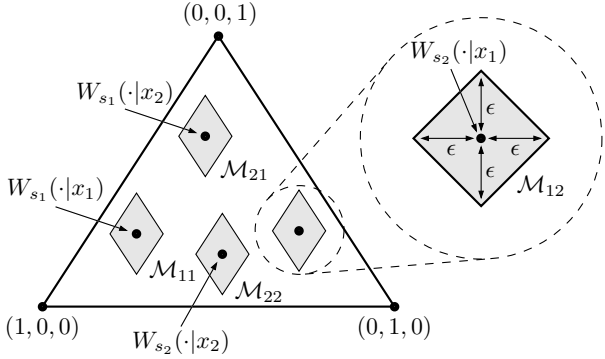

Finally, we want to illustrate this concept of distance between channels with the help of a small example. Therefore, let $\sY=\{y_1,y_2,y_3\}$, $\sX=\{x_1,x_2\}$, and $\sS=\{s_1,s_2\}$ with $|\sY|=3$, $|\sX|=2$, and $|\sS|=2$ respectively. Having two possible states, the (single-user) compound channel $\cc_1=\{W_{s_1},W_{s_2}\}$ consists of two possible channel realizations $W_{s_1}$ and $W_{s_2}$. Such a compound channel is visualized in Fig.~\ref{fig:compound}.

Now, if one is interested in constructing a second AVC $\cc_2=\{\widetilde{W}_s\}_{s\in\widetilde{\sS}}$ with state set $\widetilde{\sS}$ that is ``close'' to $\cc_1$ in the sense that $D(\cc_1,\cc_2)<\epsilon$, cf. \eqref{eq:cc_defdistance}, then the following conditions must be satisfied:
\begin{align*}
	\widetilde{W}_s(\cdot|x_1)\in\sM_1 &\qquad\text{for all }s\in\widetilde{\sS} \\
	\widetilde{W}_s(\cdot|x_2)\in\sM_2 &\qquad\text{for all }s\in\widetilde{\sS}
\end{align*}
with $\sM_1=\sM_{11}\cup\sM_{12}$ and $\sM_2=\sM_{21}\cup\sM_{22}$ the $\epsilon$-regions around the initial channel realizations as defined in Fig. \ref{fig:compound}. Note that the reverse direction holds as well. 

This discussion shows the necessity of defining the distance between compound channels in terms of $\{W_s\}_{s\in\sS}$ and $\{\widetilde{W}_s\}_{s\in\widetilde{\sS}}$ and not by the state sets $\sS$ and $\widetilde{\sS}$ themselves. Obviously, with the former definition, two compound channels can be ``close'' in their uncertainty sets although their state sets may differ a lot. For example, in the previous discussion, the constructed state set $\widetilde{\sS}$ could consist of infinitely many elements which is quite different from the original set $\sS$ which consists of only two elements.

\subsection{Continuity of Compound Secrecy Capacity}
\label{sec:cc_cont}

Now we want to study what happens if there are small variations in the uncertainty set. Obviously, it is desirable to have a \emph{continuous} behavior of the secrecy capacity meaning that small variations in the uncertainty set should only lead to small variations in the corresponding secrecy capacity. For the analysis, we need two important lemmas. Similar results were first stated and proved in \cite{Alicki04ContinuityQuantumConditionalInformation} and \cite{Leung09ContinuityQuantumCapacities} in the context of quantum information theory. In this paper, we consider classical probability distributions only which allow us to obtain similar results with better constants.

\begin{lemma}
\label{lem:cc_lem1}
Let $\sX$ and $\sY$ be finite alphabets and $\epsilon\in(0,1)$ be arbitrary. Then for all joint probability distributions $P_{XY},Q_{XY}\in\sP(\sX\times\sY)$ with $\|P_{XY}-Q_{XY}\|\leq\epsilon$ it holds that
\begin{equation}
	\big|H(Y|X\|P_{XY}) - H(Y|X\|Q_{XY})\big| \leq \delta_1(\epsilon,|\sY|) 
	\label{eq:cc_lem1}
\end{equation}
with $\delta_1(\epsilon,|\sY|) \coloneqq2\epsilon\log|\sY|+2H_2(\epsilon)$. Here, $H(Y|X\|P_{XY})$ denotes the conditional entropy of $Y$ given $X$ when $X$ and $Y$ are distributed according to the joint probability distribution $P_{XY}\in\sP(\sX\times\sY)$.
\end{lemma}
\begin{IEEEproof}
The proof is an adaptation of the corresponding proof in \cite{Alicki04ContinuityQuantumConditionalInformation} for quantum sources. However, restricting ourselves to classical probability distributions allow us to obtain better constants. For completeness, the proof can be found in Appendix \ref{sec:app_proof_lem1}.
\end{IEEEproof}
\vspace*{0.5\baselineskip}

Note that the right hand side of \eqref{eq:cc_lem1} depends only on the size of the alphabet $\sY$, but it is independent of $\sX$. This observation will be crucial for the proof of Theorem \ref{the:cc_continuity}.

\begin{lemma}
\label{lem:cc_lem2}
Let $\sX$ and $\sY$ be finite alphabets and $W,\Wtil:\sX\rightarrow\sP(\sY)$ be arbitrary channels with
\begin{equation}
	d(W,\Wtil) \leq \epsilon
	\label{eq:cc_lem2_d}
\end{equation}
for some $\epsilon>0$. For arbitrary $n\in\N$, let $\sU$ be an arbitrary finite set, $P_U\in\sP(\sU)$ the uniform distribution on $\sU$, and $E(x^n|u)$, $x^n\in\sX^n$, an arbitrary stochastic encoder, cf. \eqref{eq:cc_encoder}. We consider the probability distributions
\begin{align*}
	P_{UY^n}(u,y^n) &= \sum_{x^n\in\sX^n}W^n(y^n|x^n)E(x^n|u)P_U(u) \\
	\Ptil_{UY^n}(u,y^n) &= \sum_{x^n\in\sX^n}\Wtil^n(y^n|x^n)E(x^n|u)P_U(u).
\end{align*}
Then it holds that
\begin{equation}
	\big|I(U;Y^n\|P) - I(U;Y^n\|\Ptil)\big| \leq n\delta_2(\epsilon,|\sY|) 
	\label{eq:cc_lem2}
\end{equation}
with $\delta_2(\epsilon,|\sY|) \coloneqq 4\epsilon\log|\sY|+4H_2(\epsilon)$.
\end{lemma}
\begin{IEEEproof}
The proof is an adaptation of the proof in \cite{Leung09ContinuityQuantumCapacities} for quantum capacities. Considering classical probability distributions allows proving results with better constants. For completeness, the proof can be found in Appendix \ref{sec:app_proof_lem2}.
\end{IEEEproof}
\vspace*{0.5\baselineskip}

Note that inequality \eqref{eq:cc_lem2} depends only on the size of the output alphabet $\sY$, but is independent of the size of $\sU$ and the chosen stochastic encoder.

\begin{theorem}
\label{the:cc_continuity}
Let $\epsilon\in(0,1)$ be arbitrary. Let $\cwc_1$ and $\cwc_2$ be uncertainty sets with corresponding state sets $\sS_1$ and $\sS_2$ defining two compound wiretap channels. If 
\begin{equation*}
	D(\cwc_1,\cwc_2) < \epsilon,
\end{equation*}
then it holds that
\begin{align}
	\big|\Ccco-\Ccct\big| \leq \delta(\epsilon,|\sY|,|\sZ|) 
	\label{eq:cc_continuity}
\end{align}
with $\delta(\epsilon,|\sY|,|\sZ|)\coloneqq4\epsilon\log|\sY||\sZ| + 8H_2(\epsilon)$ a constant depending on the distance $\epsilon$ and the output alphabet sizes $|\sY|$ and $|\sZ|$.
\end{theorem}
\begin{IEEEproof}
Let $\xi>0$ be arbitrary but fixed. There exists an $\hat{s}_1=\hat{s}_1(\xi)$ such that
\begin{equation*}
	\inf_{s_1\in\sS_1}I(U;Y^n\|P^{s_1}) \geq I(U;Y^n\|P^{\hat{s}_1})-\xi.
\end{equation*}
By assumption, there exists also an $\hat{s}_2=\hat{s}_2(\hat{s}_1)$ such that $d(W_{\hat{s}_1},W_{\hat{s}_2}) < \epsilon$. This implies 
\begin{equation*}
	\big|I(U;Y^n\|P^{\hat{s}_1})-I(U;Y^n\|P^{\hat{s}_2})\big| \leq n\delta_2(\epsilon,|\sY|) 
\end{equation*}
by Lemma \ref{lem:cc_lem2}, cf. \eqref{eq:cc_lem2}. With this we obtain
\begin{align}
	\inf_{s_1\in\sS_1}I(U;Y^n\|P^{s_1}) &\geq I(U;Y^n\|P^{\hat{s}_2})-n\delta_2(\epsilon,|\sY|)-\xi \nonumber \\
	&\geq \inf_{s_2\in\sS_2}I(U;Y^n\|P^{s_2})-n\delta_2(\epsilon,|\sY|)-\xi.
	\label{eq:cc_cont_infi}
\end{align}
Note that relation \eqref{eq:cc_cont_infi} holds for all $\xi>0$. Since, the left hand side does not depend on $\delta$, we obtain
\begin{align}
	\inf_{s_1\in\sS_1}I(U;Y^n\|P^{s_1}) \geq \inf_{s_2\in\sS_2}I(U;Y^n\|P^{s_2})-n\delta_2(\epsilon,|\sY|).
	\label{eq:cc_cont_infi2}
\end{align}

We observe that if we exchange the roles of $\sS_1$ and $\sS_2$ in the previous derivation, we end up with an expression as in \eqref{eq:cc_cont_infi2}, where the infima over $\sS_1$ and $\sS_2$ are interchanged. Accordingly, this means that
\begin{equation*}
	\big|\inf_{s_1\in\sS_1}I(U;Y^n\|P^{s_1})-\inf_{s_2\in\sS_2}I(U;Y^n\|P^{s_2})\big| \leq n\delta_2(\epsilon,|\sY|). 
\end{equation*}
The same arguments lead to
\begin{equation*}
	\big|\sup_{s_1\in\sS_1}I(U;Z^n\|P^{s_1})-\sup_{s_2\in\sS_2}I(U;Z^n\|P^{s_2})\big| \leq n\delta_2(\epsilon,|\sZ|) 
\end{equation*}
so that we conclude 
\begin{align*}
	&\Big|\inf_{s_1\in\sS_1}I(U;Y^n\|P^{s_1})-\sup_{s_1\in\sS_1}I(U;Z^n\|P^{s_1}) \\
	&\qquad\qquad- \big(\inf_{s_2\in\sS_2}I(U;Y^n\|P^{s_2})-\sup_{s_2\in\sS_2}I(U;Z^n\|P^{s_2})\big)\Big| \\
	&\qquad\leq \Big|\inf_{s_1\in\sS_1}I(U;Y^n\|P^{s_1})-\inf_{s_2\in\sS_2}I(U;Y^n\|P^{s_2})\Big| \\
	&\qquad\qquad + \Big|\sup_{s_1\in\sS_1}I(U;Z^n\|P^{s_1})-\sup_{s_2\in\sS_2}I(U;Z^n\|P^{s_2})\Big| \\
	&\qquad\leq n\delta_2(\epsilon,|\sY|) + n\delta_2(\epsilon,|\sZ|) = n\delta(\epsilon,|\sY|,|\sZ|) 
\end{align*}
with $\delta(\epsilon,|\sY|,|\sZ|)=4\epsilon\log|\sY||\sZ|+ 8H_2(\epsilon)$. But this implies for the secrecy capacities
\begin{align*}
	&\frac{1}{n}\Big(\inf_{s_1\in\sS_1}I(U;Y^n\|P^{s_1})-\sup_{s_1\in\sS_1}I(U;Z^n\|P^{s_1})\Big) \\
	&\quad \leq \frac{1}{n}\Big(\inf_{s_2\in\sS_2}I(U;Y^n\|P^{s_2})-\sup_{s_2\in\sS_2}I(U;Z^n\|P^{s_2})\Big) \\
	&\quad\qquad\qquad+ \delta(\epsilon,|\sY|,|\sZ|)
\end{align*}
so that
\begin{equation}
	\Ccco \leq \Ccct + \delta(\epsilon,|\sY|,|\sZ|).
	\label{eq:cc_cont_secrecy}
\end{equation}
Again, we can exchange the roles of $\sS_1$ and $\sS_2$ in the derivation above to obtain a relation as in \eqref{eq:cc_cont_secrecy} where $\Ccco$ and $\Ccct$ are interchanged. Thus, we have
\begin{align*}
	\big|\Ccco-\Ccct\big| \leq \delta(\epsilon,|\sY|,|\sZ|) 
\end{align*}
which proves the desired result.
\end{IEEEproof}

\begin{remark}
\label{rem:quant}
Note that \eqref{eq:cc_continuity} explicitly quantifies by $\delta(\epsilon,|\sY|,|\sZ|)$ how much the secrecy capacity can differ in terms of the distance $\epsilon$ and the channel output alphabets $|\sY|$ and $|\sZ|$.
\end{remark}

\begin{corollary}
\label{cor:cc_limit}
For any compound wiretap channel with uncertainty set $\cwc$ and any sequence $\{\cwc_n\}_{n\in\N}$ satisfying $\lim_{n\rightarrow\infty}D(\cwc,\cwc_n)=0$, it holds that
\begin{equation*}
	\Ccc = \lim_{n\rightarrow\infty}\Cccn.
\end{equation*}
\end{corollary}
\begin{IEEEproof}
The result follows immediately from \eqref{eq:cc_continuity} of Theorem \ref{the:cc_continuity}.
\end{IEEEproof}

\subsection{Discussion}
\label{sec:cc_discussion}

In this section we have shown that the secrecy capacity of the compound wiretap channel is a continuous function of the uncertainty set. This means that small variations in the uncertainty set result only in small variations of the corresponding secrecy capacity. This is in particular crucial in the context of adversaries, where the uncertainty set reflects the adversary's strategy space. It is a necessary requirement for a system design to be robust, i.e., having continuous dependency, against such changes in strategies.

Finally, we want to highlight that the continuity of the secrecy capacity was established without having a single-letter description available. Although a multi-letter characterization of the secrecy capacity as given in Theorem \ref{the:cc_secrecy} is not efficiently computable, it is extremely useful for deriving certain properties such as continuity as demonstrated in Theorem \ref{the:cc_continuity}.

\section{Arbitrarily Varying Wiretap Channel}
\label{sec:avc}

\begin{figure}
  \centering
  	\scalebox{0.9}{\hspace{-2.75cm}
  	\begin{picture}(0,0)%
  	\includegraphics{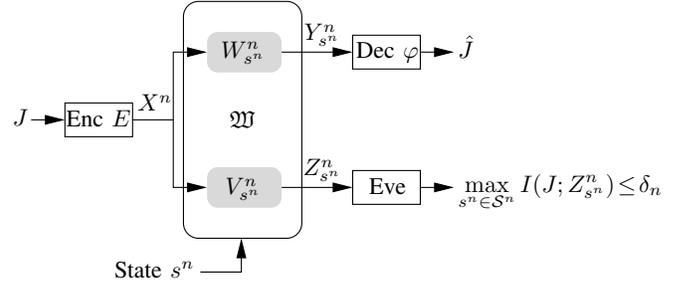}%
  	\end{picture}%
  	\setlength{\unitlength}{3947sp}%
  	\begingroup\makeatletter\ifx\SetFigFont\undefined%
  	\gdef\SetFigFont#1#2#3#4#5{%
  	  \fontfamily{#3}\fontseries{#4}\fontshape{#5}%
  	  \selectfont}%
  	\fi\endgroup%
  	\begin{picture}(3031,1996)(2800,-2208)
  	\put(5816,-1554){\makebox(0,0)[lb]{\smash{{\SetFigFont{8}{9.6}{\rmdefault}{\mddefault}{\updefault}{\color[rgb]{0,0,0}$\displaystyle\max_{s^n\in\sS^n}I(J;Z^n_{s^n})\!\leq\!\delta_n$}%
  	}}}}
  	\put(5816,-618){\makebox(0,0)[lb]{\smash{{\SetFigFont{8}{9.6}{\rmdefault}{\mddefault}{\updefault}{\color[rgb]{0,0,0}$\hat{J}$}%
  	}}}}
  	\put(3976,-2151){\makebox(0,0)[rb]{\smash{{\SetFigFont{8}{9.6}{\rmdefault}{\mddefault}{\updefault}{\color[rgb]{0,0,0}$\text{State } s^n$}%
  	}}}}
  	\put(2815,-1087){\makebox(0,0)[rb]{\smash{{\SetFigFont{8}{9.6}{\rmdefault}{\mddefault}{\updefault}{\color[rgb]{0,0,0}$J$}%
  	}}}}
  	\put(3700,-988){\makebox(0,0)[b]{\smash{{\SetFigFont{8}{9.6}{\rmdefault}{\mddefault}{\updefault}{\color[rgb]{0,0,0}$X^n$}%
  	}}}}
  	\put(4868,-495){\makebox(0,0)[b]{\smash{{\SetFigFont{8}{9.6}{\rmdefault}{\mddefault}{\updefault}{\color[rgb]{0,0,0}$Y^n_{s^n}$}%
  	}}}}
  	\put(4868,-1440){\makebox(0,0)[b]{\smash{{\SetFigFont{8}{9.6}{\rmdefault}{\mddefault}{\updefault}{\color[rgb]{0,0,0}$Z^n_{s^n}$}%
  	}}}}
  	\put(5324,-616){\makebox(0,0)[b]{\smash{{\SetFigFont{8}{9.6}{\rmdefault}{\mddefault}{\updefault}{\color[rgb]{0,0,0}$\text{Dec } \varphi$}%
  	}}}}
  	\put(5324,-1567){\makebox(0,0)[b]{\smash{{\SetFigFont{8}{9.6}{\rmdefault}{\mddefault}{\updefault}{\color[rgb]{0,0,0}$\text{Eve}$}%
  	}}}}
  	\put(3308,-1100){\makebox(0,0)[b]{\smash{{\SetFigFont{8}{9.6}{\rmdefault}{\mddefault}{\updefault}{\color[rgb]{0,0,0}$\text{Enc } E$}%
  	}}}}
  	\put(4312,-1104){\makebox(0,0)[b]{\smash{{\SetFigFont{8}{9.6}{\rmdefault}{\mddefault}{\updefault}{\color[rgb]{0,0,0}$\avwc$}%
  	}}}}
  	\put(4316,-1572){\makebox(0,0)[b]{\smash{{\SetFigFont{8}{9.6}{\rmdefault}{\mddefault}{\updefault}{\color[rgb]{0,0,0}$V_{s^n}^n$}%
  	}}}}
  	\put(4316,-618){\makebox(0,0)[b]{\smash{{\SetFigFont{8}{9.6}{\rmdefault}{\mddefault}{\updefault}{\color[rgb]{0,0,0}$W_{s^n}^n$}%
  	}}}}
  	\end{picture}%
  	}
  \caption{Arbitrarily varying wiretap channel. In contrast to the compound wiretap channel, the transmission is now governed by an unknown state sequence $s^n\in\sS^n$ of length $n$, which may vary in an unknown manner from channel use to channel use.}
  \label{fig:avc}
\end{figure}

We continue our analysis with the \emph{arbitrarily varying wiretap channel}. In contrast to the previously studied compound wiretap channel, the unknown channel realization may vary in an unknown and arbitrary manner from channel use to channel use. The AVWC is depicted in Fig.~\ref{fig:avc}.

This model includes the scenario of active adversaries, who are able to jam the legitimate transmission. Since an adversary will not share any information about the jamming signal, the legitimate users must be prepared for the worst: arbitrary and unknown interfering sequences. This motivates the use of AVCs for such scenarios as well.

The main result for the AVWC will be that its unassisted secrecy capacity is a discontinuous function of the uncertainty set. Accordingly, unassisted strategies will not be robust against such classes of attacks. In particular, we will show that even for the simplest case of an uncertainty set with only two states, the corresponding unassisted secrecy capacity will be discontinuous in the uncertainty set. Obviously, this is then also true for more general AVWCs, whose uncertainty sets contain more elements. Note that the case of an uncertainty set with one element does not define an AVWC (since the channel realization naturally stays constant for the whole duration of transmission), but belongs to compound wiretap channels, whose secrecy capacity is continuous.

As before let $\sX$, $\sY$, and $\sZ$ be finite input and output sets and $\sS$ be the state set. But in contrast to the compound wiretap channel in Section \ref{sec:cc}, we assume $\sS$ to be finite in the sequel. Then for a fixed state sequence $s^n\in\sS^n$ of length $n$, the discrete memoryless channel to the legitimate receiver is given by $W_{s^n}^n(y^n|x^n)=W^n(y^n|x^n,s^n)\coloneqq\prod_{i=1}^nW(y_i|x_i,s_i)$. 

Then the family of channels for all $s^n\in\sS^n$ defines the (marginal) AVC to the legitimate receiver as
\begin{equation*}
	\avc \coloneqq \big\{W_{s^n}^n\big\}_{s^n\in\sS^n}.
\end{equation*}
In addition, for any probability distribution $q\in\sP(\sS)$ we define the \emph{averaged channel} as
\begin{equation}
	W_q(y|x) = \sum_{s\in\sS}W(y|x,s)q(s).
	\label{eq:avc_averaged}
\end{equation}

An important property of an AVC is the so-called concept of symmetrizability as defined below. 

\begin{definition}
\label{def:avc_symmetrizable}
An AVC is called \emph{symmetrizable} if there exists a stochastic matrix $\sigma:\sX\rightarrow \sP(\sS)$ such that
\begin{equation*}
	\sum_{s\in\sS}W(y|x_1,s)\sigma(s|x_2) = \sum_{s\in\sS}W(y|x_2,s)\sigma(s|x_1)
\end{equation*}
holds for all $x_1,x_2\in\sX$ and $y\in\sY$.
\end{definition}

Roughly speaking, a symmetrizable AVC can ``simulate'' a valid input, which makes it impossible for the decoder to decide on the correct codeword.

Similarly for the channel to the eavesdropper, we define the discrete memoryless channel as $V_{s^n}^n(z^n|x^n)=V^n(z^n|x^n,s^n)\coloneqq\prod_{i=1}^nV(z_i|x_i,s_i)$ for a given state sequence $s^n\in\sS^n$. Further, we set $\mathcal{V}\coloneqq\{V_{s^n}^n\}_{s^n\in\sS^n}$ and $V_q(z|x) = \sum_{s\in\sS}V(z|x,s)q(s)$ for $q\in\sP(\sS)$.

\begin{definition}
\label{def:avc_avwc}
The discrete memoryless \emph{arbitrarily varying wiretap channel} is given by the families of pairs of channels with common input as
\begin{equation*}
	\avwc \coloneqq \big\{\avc,\avcv\big\}=\big\{W_{s^n}^n,V_{s^n}^n\big\}_{W_{s^n}^n\in\avc,V_{s^n}^n\in\avcv}. 
\end{equation*}
\end{definition}

\subsection{Unassisted Codes}
\label{sec:avc_det}

The definition of an \emph{unassisted $(n,J_n)$-code} $\Cdet$ for the AVWC is the same as for the compound wiretap channel in Definition \ref{def:cc_code}: It consists of a stochastic encoder as in \eqref{eq:cc_encoder} and a deterministic decoder as in \eqref{eq:cc_decoder}.

The difference lies in the reliability and secrecy criteria as we have now to consider state sequences $s^n\in\sS^n$ of length $n$. Thus, for given $s^n\in\sS^n$ the average probability of decoding error at the legitimate receiver is
\begin{align*}
	\bar{e}_n(s^n\|\Cdet) \coloneqq  \frac{1}{|\sJ_n|}\sum_{j\in\sJ_n}\sum_{x^n\in\sX^n}\sum_{y^n:\varphi(y^n)\neq j}\!\!\!W^n(y^n|x^n,s^n)E(x^n|j)
\end{align*}
and the confidentiality of the message is measured by $\max_{s^n\in\sS^n}I(J;Z_{s^n}^n\|\Cdet)\leq\delta_n$ with $Z_{s^n}^n=(Z_{s_1},Z_{s_2},...,Z_{s_n})$.

\begin{definition}
\label{def:avc_detachievable}
A rate $R>0$ is said to be an \emph{achievable secrecy rate} for the AVWC if for any $\tau>0$ there exist an $n(\tau)\in\N$, positive null sequences $\{\lambda_n\}_{n\in\N}$, $\{\delta_n\}_{n\in\N}$, and a sequence of $(n,J_n)$-codes $\{\Cdet_n\}_{n\in\N}$ such that for all $n\geq n(\tau)$ we have $\frac{1}{n}\log J_n\geq R-\tau$,
\begin{align}
	\max_{s^n\in\sS^n}\bar{e}_n(s^n\|\Cdet_n)&\leq\lambda_n,
	\label{eq:avc_achievable1}
\intertext{and}
	\max_{s^n\in\sS^n}I(J;Z_{s^n}^n\|\Cdet_n)&\leq\delta_n.
	\label{eq:avc_achievable2}
\end{align} 
The \emph{unassisted secrecy capacity} $\Cavc$ of the AVWC with uncertainty set $\avwc$ is given by the supremum of all achievable secrecy rates $R$.
\end{definition} 

\begin{remark}
\label{rem:strategies}
Conditions \eqref{eq:avc_achievable1} and \eqref{eq:avc_achievable2} already indicate that an active adversary may have different strategies. On the one hand, it can try to disturb the legitimate communication as much as possible by choosing the state sequence in such a way that the probability of error \eqref{eq:avc_achievable1} is maximized. On the other hand, it can try to maximize the information leakage \eqref{eq:avc_achievable2}. Of course, any strategy in between is also a valid jamming strategy.
\end{remark}

Unfortunately, such unassisted approaches do not suffice to establish reliable communication over \emph{symmetrizable} AVCs; indeed, the corresponding capacity is zero in this case \cite{CsiszarNarayan88AVCRevisited,Bjelakovic13AVWiretap,BocheSchaefer13Superactivation}. This necessitates the use of more sophisticated strategies based on \emph{common randomness}.

\subsection{CR-Assisted Codes}
\label{sec:avc_ran}

CR is modeled by a random variable $\Gamma$ taking values in a finite set $\sG_n$ according to a distribution $P_\Gamma\in\sP(\sG_n)$. It enables transmitter and receiver to coordinate their choices of encoder \eqref{eq:cc_encoder} and decoder \eqref{eq:cc_decoder} according to the realization $\gamma\in\sG_n$.

\begin{definition}
\label{def:avc_rancode}
A \emph{CR-assisted} $(n,J_n,\sG_n,P_\Gamma)$-\emph{code} $\Cran$ is given by a family of unassisted codes
\begin{equation*}
	\big\{\Cdet(\gamma):\gamma\in\sG_n\big\}
\end{equation*}
together with a random variable $\Gamma$ taking values in $\sG_n$ with $|\sG_n|<\infty$ according to $P_\Gamma\in\sP(\sG_n)$.
\end{definition}

The reliability and secrecy constraints from above extend to CR-assisted codes in the following way: The mean average probability of error for $s^n\in\sS^n$ is given by $\bar{e}_{\text{CR},n}(s^n\|\Cran)=\E_\Gamma[\bar{e}_n(s^n\|\Cdet(\Gamma))]$, i.e.,
\begin{align*}
	&\bar{e}_{\text{CR},n}(s^n\|\Cran)\coloneqq \frac{1}{|\sJ_n|}\sum_{j\in\sJ_n}\sum_{\gamma\in\sG_n}\sum_{x^n\in\sX^n} \\
	&\qquad\qquad \times \sum_{y^n:\varphi_\gamma(y^n)\neq j}W^n(y^n|x^n,s^n)E_\gamma(x^n|j)P_\Gamma(\gamma).
\end{align*}
With $I(J;Z_{s^n}^n\|\Cran)=\E_\Gamma[I(J;Z_{s^n}^n\|\Cdet(\Gamma))]$ the secrecy requirement becomes
\begin{equation*}
	\max_{s^n\in\sS^n}\sum_{\gamma\in\sG_n}I(J;Z_{s^n}^n\|\Cdet(\gamma))P_\Gamma(\gamma)\leq\delta_n.
\end{equation*}
Definitions of a \emph{CR-assisted achievable secrecy rate} and the \emph{CR-assisted secrecy capacity} $\Cavccr$ follow accordingly.

\subsection{Capacity Results}
\label{sec:avc_capacity}

Studies have considered the secrecy capacity of the AVWC \cite{MolavianJazi09ArbitraryJamming,Bjelakovic13AVWiretap,BocheSchaefer13Superactivation,BocheSchaeferPoor14AVWCSecrecyMeasures,WieseXXAVWC,NoetzelXXAVWC} where the latter use the strong secrecy criterion. In particular, the relation between the secrecy capacities for unassisted and CR-assisted codes has been completely characterized in \cite{NoetzelXXAVWC}. 

\begin{theorem}[\cite{NoetzelXXAVWC}]
\label{the:avc_detcapacity}
If the CR-assisted secrecy capacity satisfies $\Cavccr>0$, then the unassisted secrecy capacity is given by
\begin{equation*}
	\Cavc = \Cavccr
\end{equation*}
if and only if the AVC $\avc$ to the legitimate receiver is non-symmetrizable. If the AVC $\avc$ is symmetrizable, then $\Cavc=0$. If $\Cavc=0$ and $\Cavccr>0$, then the AVC $\avc$ is symmetrizable.
\end{theorem}

The unassisted secrecy capacity $\Cavc$ of the AVWC $\avwc$ is completely known in terms of its CR-assisted secrecy capacity $\Cavccr$. A multi-letter description of $\Cavccr$ itself has been recently established in \cite{WieseXXAVWC}, while a single-letter expression remains open.

\begin{remark}
\label{rem:continuity}
The multi-letter description of the CR-assisted secrecy capacity $\Cavccr$ can now be used to show that it depends in a continuous way on the corresponding uncertainty set $\avwc$. This can be done as in Section~\ref{sec:cc_codes} for the compound wiretap channel, where the multi-letter description of the secrecy capacity $\Ccc$ in Theorem \ref{the:cc_secrecy} is used in Theorem~\ref{the:cc_continuity} to show that it depends continuously on $\cwc$.
\end{remark}

\section{Discontinuity of AVWC Secrecy Capacity}
\label{sec:avc_continuity}

In this section we study the continuity of the unassisted secrecy capacity $\Cavc$ of the AVWC with uncertainty set $\avwc$. We will use Theorem \ref{the:avc_detcapacity} which provides a characterization in terms of its corresponding CR-assisted secrecy capacity $\Cavccr$. But in contrast to the previously studied compound wiretap channel, we do not make use of any multi-letter characterization of the secrecy capacity as in Section \ref{sec:cc_continuity}.

Nonetheless, we will be able to show by elementary calculations that the unassisted secrecy capacity $\Cavc$ is discontinuous in the uncertainty set $\avwc$. Similar to the compound wiretap channel, we ask the question: if the distance between two AVWCs is small, i.e., $D(\avwc_1,\avwc_2)< \epsilon$, does this imply that $\Cavco-\Cavct$ is small as well?

In more detail, let $\{\avwc_n\}_{n\in\N}$ be a sequence of finite uncertainty sets, which converges to a finite set $\avwc^*$ in terms of $D$-distance. The question is then whether or not this implies
\begin{equation}
	\lim_{n\rightarrow\infty}\Cavcn = \Cavcs.
	\label{eq:avc_continuity}
\end{equation}
In the following we will examine this equation via a simple example to show that \eqref{eq:avc_continuity} does not hold in general.

\subsection{Secrecy Capacity with Discontinuity Point}
\label{sec:avc_continuity_point}

The aim of this part is to construct an AVWC whose unassisted secrecy capacity has a discontinuity point. To do so, we consider a communication scenario with input and output alphabets of sizes $|\sX|=2$, $|\sY|=3$, $|\sZ|=2$, and $|\sS|=2$.

Let us first consider the communication channel to the legitimate receiver. To construct a suitable AVC for this link, we make use of an example which first appeared in \cite{Blackwell60AVC} and which was later also discussed in \cite[Example 1]{Ahlswede78EliminationCorrelationAVC}. We follow this example and construct an AVC to the legitimate receiver with uncertainty set
\begin{equation}
	\avc=\big\{W_1,W_2\big\}
	\label{eq:avc_avc}
\end{equation}	
where
\begin{equation*}
	W_1 \coloneqq \begin{pmatrix} 1 & 0 & 0 \\ 0 & 0 & 1 \end{pmatrix} \quad\text{and}\quad W_2 \coloneqq \begin{pmatrix} 0 & 0 & 1 \\ 0 & 1 & 0 \end{pmatrix}.
\end{equation*}
We know from \cite{Ahlswede78EliminationCorrelationAVC} that $\avc$ defines a symmetrizable AVC so that its unassisted capacity is zero, i.e., $C(\avc)=0$. 

Further, with the channel
\begin{equation}
	\hat{W} \coloneqq \begin{pmatrix} 1 & 0 & 0 \\ 0 & 1 & 0 \end{pmatrix}
	\label{eq:avc_what}
\end{equation}
we define the trivial AVC whose two elements are identical as
\begin{equation}
	\hat{\avc}=\big\{\hat{W},\hat{W}\big\}.
	\label{eq:avc_avchat}
\end{equation}

Now, for the channel to the eavesdropper, we define the ``useless'' channel
\begin{equation}
	V \coloneqq \begin{pmatrix} \frac{1}{2} & \frac{1}{2}  \\ \frac{1}{2} & \frac{1}{2} \end{pmatrix}.
	\label{eq:avc_v}
\end{equation}
Then the set $\avcv=\{V,V\}$ defines a corresponding AVC to the eavesdropper.

These definitions finally create with \eqref{eq:avc_avc}, \eqref{eq:avc_avchat}, and \eqref{eq:avc_v} the following two AVWCs specified by their uncertainty sets:
\begin{align*}
	\avwc \coloneqq \big\{\avc,\avcv\big\} \quad\text{and}\quad
	\hat{\avwc} \coloneqq \big\{\hat{\avc},\avcv\big\}.
\end{align*}
Moreover, we can define a convex combination of these two AVWCs as
\begin{equation}
	\avwc(\lambda) = \big\{\{W_1(\lambda),W_2(\lambda)\},\avcv\big\} \quad\text{for }0\leq\lambda\leq1,
	\label{eq:avc_avwcconvex}
\end{equation}
with convex combinations
\begin{subequations}
\label{eq:avc_convexcombinations}
\begin{align}
	W_1(\lambda)&=W_{1,\lambda}=(1-\lambda)W_1+\lambda \hat{W} \\
	W_2(\lambda)&=W_{2,\lambda}=(1-\lambda)W_2+\lambda \hat{W}.
\end{align}
\end{subequations}
Note that \eqref{eq:avc_avwcconvex} is indeed a convex combination of the eavesdropper AVC as well which is trivial in this case as we have identical elements.

Now, the following result shows that the unassisted secrecy capacity $C_S(\avwc(\lambda))$ is discontinuous in $\lambda$.

\begin{theorem}
\label{the:avc_discont}
The following assertions hold for the previous example:
\begin{enumerate}
	\item The CR-assisted secrecy capacity $\Cavccrlambda$ is continuous in $\lambda$ for all $\lambda\in[0,1]$ and it holds that
	\begin{equation}
		\min_{\lambda\in[0,1]}\Cavccrlambda>0.
		\label{eq:avc_theorem1}
	\end{equation}
	\item The unassisted secrecy capacity $C_S(\avwc(\lambda))$ is continuous in $\lambda$ for all $\lambda\in(0,1]$. It holds that $C_S(\avwc(0))=0$ and further that
	\begin{equation}
		\lim_{\lambda\searrow0}C_S(\avwc(\lambda))>0,
		\label{eq:avc_theorem2}
	\end{equation}
	i.e., $\lambda=0$ is a discontinuity point of $C_S(\cdot)$.
\end{enumerate}
\end{theorem}
\begin{IEEEproof}
Let us first give an outline of the proof which is divided into several steps. First, we will show that the AVC $\avc(\lambda)=\{W_1(\lambda),W_2(\lambda)\}$ is non-symmetrizable for all $\lambda\in(0,1]$. This implies then that
\begin{equation}
	C_S(\avwc(\lambda)) = \Cavccrlambda \quad\text{for all }\lambda\in(0,1]
	\label{eq:avc_cequal}
\end{equation}
by Theorem \ref{the:avc_detcapacity}. As a second step, we will then show that $\Cavccrlambda$ is continuous for all $\lambda\in[0,1]$ and further
\begin{equation}
	\min_{\lambda\in[0,1]}\Cavccrlambda>0.
	\label{eq:avc_mincr}
\end{equation}
On the other hand, since $\avc=\{W_1,W_2\}$ is symmetrizable, we have
\begin{equation*}
	C_S(\avwc(0)) = C_S(\avwc) = 0.
\end{equation*}
But due to \eqref{eq:avc_mincr}, $C_S(\avwc(\lambda))$ is discontinuous in $\lambda=0$. This will then conclude the proof.

Before we start proving the first step, we define suitable channels given by matrices
\begin{align*}
	\vec{W}_1 = \begin{pmatrix}
	W_{11}^T \\ W_{12}^T
	\end{pmatrix}
	\quad\text{and}\quad
	\vec{W}_2 = \begin{pmatrix}
		W_{21}^T \\ W_{22}^T
	\end{pmatrix}
\end{align*}
with vectors
\begin{gather*}
	W_{11} = \begin{pmatrix} 1 \\ 0 \\ 0 \end{pmatrix},\; W_{12} = \begin{pmatrix} 0 \\ 0 \\ 1 \end{pmatrix}, \;
	W_{21} = \begin{pmatrix} 0 \\ 0 \\ 1 \end{pmatrix},\; W_{22} = \begin{pmatrix} 0 \\ 1 \\ 0 \end{pmatrix}.
\end{gather*}
Similarly as in \eqref{eq:avc_convexcombinations}, we define convex combinations of these vectors with the rows of $\hat{W}$ in \eqref{eq:avc_what}. In more detail, for each $\lambda\in(0,1)$ we have
\begin{align*}
	W_{11}(\lambda) &= (1-\lambda)\begin{pmatrix}1\\0\\0\end{pmatrix} + \lambda\begin{pmatrix}1\\0\\0\end{pmatrix} = \begin{pmatrix}1\\0\\0\end{pmatrix} \\
	W_{12}(\lambda) &= (1-\lambda)\begin{pmatrix}0\\0\\1\end{pmatrix} + \lambda\begin{pmatrix}0\\1\\0\end{pmatrix} = \begin{pmatrix}0\\\lambda\\1-\lambda\end{pmatrix} \\
	W_{21}(\lambda) &= (1-\lambda)\begin{pmatrix}0\\0\\1\end{pmatrix} + \lambda\begin{pmatrix}1\\0\\0\end{pmatrix} = \begin{pmatrix}\lambda\\0\\1-\lambda\end{pmatrix} \\
	W_{22}(\lambda) &= (1-\lambda)\begin{pmatrix}0\\1\\0\end{pmatrix} + \lambda\begin{pmatrix}0\\1\\0\end{pmatrix} = \begin{pmatrix}0\\1\\0\end{pmatrix}.
\end{align*}

Now we are in the position to show the first step as outlined above. Therefore, we first show by contradiction that the AVC $\avc(\lambda)=\{W_1(\lambda),W_2(\lambda)\}=\{W_{s,\lambda}\}_{s=1,2}$ is non-symmetrizable for all $\lambda\in(0,1]$. Accordingly, we assume that this channel is symmetrizable, i.e., for each $\lambda\in(0,1]$ there exists a stochastic matrix $\sigma:\sX\rightarrow\sP(\sS)$ with 
\begin{equation*}
	\sum_{s\in\sS}W_{s,\lambda}(y|x_1)\sigma(s|x_2) = \sum_{s\in\sS}W_{s,\lambda}(y|x_2)\sigma(s|x_1)
\end{equation*}
for all $y\in\{1,2,3\}$ and $x_1,x_2\in\{1,2\}$, cf. Definition \ref{def:avc_symmetrizable}. With $s\in\{1,2\}$ this would imply
\begin{equation*}
	\begin{pmatrix}1\\0\\0\end{pmatrix}\sigma(1|2)+\begin{pmatrix}\lambda\\0\\1-\lambda\end{pmatrix}\sigma(2|2) = \begin{pmatrix}0\\\lambda\\1-\lambda\end{pmatrix}\sigma(1|1)+\begin{pmatrix}0\\1\\0\end{pmatrix}\sigma(2|1).
\end{equation*}
With $\sigma(1|2)=a$, $\sigma(2|2)=1-a$, $a\in[0,1]$, and $\sigma(1|1)=b$, $\sigma(2|1)=1-b$, $b\in[0,1]$, this can be written as
\begin{equation*}
	\begin{pmatrix}1\\0\\0\end{pmatrix}a+\begin{pmatrix}\lambda\\0\\1-\lambda\end{pmatrix}(1-a) = \begin{pmatrix}0\\\lambda\\1-\lambda\end{pmatrix}b+\begin{pmatrix}0\\1\\0\end{pmatrix}(1-b).
\end{equation*}
In particular, the second line is $0 = \lambda b + 1-b$ or equivalently $b = \frac{1}{1-\lambda}$. Now, $\lambda\in(0,1]$ implies that $b>1$ which is a contradiction. Thus, for $\lambda\in(0,1]$ the AVC $\avc(\lambda)$ is non-symmetrizable and therewith $C_S(\avwc(\lambda)) = \Cavccrlambda$ for all $\lambda\in(0,1]$, cf. \eqref{eq:avc_cequal}.

Next, we prove the second step, where we want to understand the behavior of the CR-assisted secrecy capacity $\Cavccrlambda$ for all $\lambda\in[0,1]$. To do so, we look at the corresponding CR-assisted capacity of the legitimate link (no secrecy at this point). For any $q\in\sP(\sS)$, let
\begin{equation*}
	W_{q,\lambda}(y|x) = \sum_{s\in\sS}W_{s,\lambda}(y|x)q(s)
\end{equation*}
be the averaged channel as in \eqref{eq:avc_averaged}. Then we know from results for the classical AVC, cf.  \cite{Blackwell60AVC,Ahlswede78EliminationCorrelationAVC,CsiszarNarayan88AVCRevisited}, that
\begin{align}
	C_{\text{CR}}(\avc(\lambda)) &= \max_{p\in\sP(\sX)}\min_{q\in\sP(\sS)}I(p,W_{q,\lambda}) \nonumber \\
	&= \min_{q\in\sP(\sS)}\max_{p\in\sP(\sX)}I(p,W_{q,\lambda})
	\label{eq:avc_cont_cran}
\end{align}
since $\sP(\sX)$ and $\sP(\sS)$ are convex sets and the mutual information $I$ is concave in $p$ and convex in $W_{q,\lambda}$. Moreover, \eqref{eq:avc_cont_cran} is continuous in $\lambda\in[0,1]$.

Assume that there exists a $\lambda_0\in[0,1]$ with $C_{\text{CR}}(\avc(\lambda_0)) = 0$; then there must be a $q_0\in\sP(\sS)$ such that $\max_{p\in\sP(\sX)}I(p,W_{q_0,\lambda_0})=0$. This would imply for the matrix
\begin{equation*}
W_{q_0,\lambda_0}=(1-q_0)W_{1,\lambda_0}+ q_0W_{2,\lambda_0}
\end{equation*}
that both of its rows are equal, i.e., its output distribution does not depend on the particular input $x=1$ or $x=2$. And this means that
\begin{equation*}
	(1-q_0)W_{11}(\lambda_0)+q_0W_{21}(\lambda_0) = (1-q_0)W_{12}(\lambda_0)+q_0W_{22}(\lambda_0)
\end{equation*}
or equivalently
\begin{equation*}
	\begin{pmatrix}1-q_0\\0\\0\end{pmatrix} + \begin{pmatrix}q_0\lambda_0\\0\\q_0(1-\lambda_0)\end{pmatrix} = \begin{pmatrix}0\\(1-q_0)\lambda_0\\(1-q_0)(1-\lambda_0)\end{pmatrix}+\begin{pmatrix}0\\q_0\\0\end{pmatrix}.
\end{equation*}
This implies that $1-q_0 + q_0\lambda_0 = 0$ must hold, i.e.,
\begin{equation*}
	q_0=\frac{1}{1-\lambda_0}.
\end{equation*}
For $\lambda_0\in(0,1]$, again, this would imply $q_0>1$ which is impossible. For $\lambda_0=0$ on the other hand, we get $q_0=1$ so that the channel is simply given by the matrix $W_2$. But this is also not possible, since the rows are not the same. Thus, for all $\lambda\in[0,1]$ we have $C_{\text{CR}}(\avc(\lambda)) > 0$. Since the function is continuous in $\lambda$, we have
\begin{equation*}
	\min_{\lambda\in[0,1]}C_{\text{CR}}(\avc(\lambda)) > 0.
\end{equation*}
It remains to show that
\begin{equation*}
	\Cavccrlambda = C_{\text{CR}}(\avc(\lambda))
\end{equation*}
holds. Obviously, the inequality $\Cavccrlambda \leq C_{\text{CR}}(\avc(\lambda))$ holds so that we only have to show the other direction. Therefore, let $R>0$ be a CR-assisted achievable rate for non-secret communication, i.e., for each $\tau>0$ there exists an $n(\tau)\in\N$ and a sequence of $(n,J_n,\sG_n,P_\Gamma)$-codes $\Cran$ such that for all $n\geq n(\tau)$ we have $\frac{1}{n}\log|\sJ_n|\geq R-\tau$ and $\max_{s^n\in\sS^n}\bar{e}_n(s^n\|\Cran)\rightarrow0$ as $n\rightarrow\infty$. Now we have to analyze the secrecy condition $I(J;Z_{s^n}^n\|\Cran)$ for this code. The corresponding joint probability distribution is given by
\begin{equation*}
	P_{JZ_{s^n}^n}(j,z^n) = \prod_{i=1}^nV(z_i|x_i,s_i)E(x^n|j)\frac{1}{|\sJ_n|}.
\end{equation*}
Since we only have one possible channel realization $V$, cf. \eqref{eq:avc_v}, in this particular example, we have $V(z|x,s)=V(z|x)$. Moreover, since $V$ is the ``useless'' channel, the output $z^n$ in $\prod_{i=1}^nV(z_i|x_i))$ is independent of the input $x^n$ so that $I(J;Z_{s^n}^n\|\Cran) = 0$. We conclude that this rate then is also achievable under the strong secrecy criterion which then shows $\Cavccrlambda = C_{\text{CR}}(\avc(\lambda))$ so that \eqref{eq:avc_theorem1} and therewith assertion 1) of the theorem is proved.

Finally, we complete the proof for assertion 2) of the theorem. Since the AVC $\avc(\lambda)$ for $\lambda\in(0,1]$ is non-symmetrizable so that $\Cavccrlambda = C_{\text{CR}}(\avc(\lambda))$, we also have
\begin{equation*}
	\inf_{\lambda\in(0,1]}C_S(\avwc(\lambda)) = \min_{\lambda\in[0,1]}C_{\text{CR}}(\avc(\lambda))>0.
\end{equation*}
But since for $\lambda=0$ the AVC $\avc(\lambda)$ is symmetrizable, we must have $C_{S}(\avwc(0))=0$ and therewith
\begin{equation*}
	\liminf_{\lambda\rightarrow0}C_S(\avwc(\lambda)) > C_S(\avwc(0)),
\end{equation*}
i.e., $C_S(\cdot)$ is a discontinuous function which proves \eqref{eq:avc_theorem2} and therewith the assertion 2) of the theorem.
\end{IEEEproof}

\begin{remark}
\label{rem:avc_weak}
Note that Theorem \ref{the:avc_discont} is proven for the strong secrecy criterion. However, it also holds for the weak secrecy criterion, which means that weakening the secrecy requirement does not help to overcome the discontinuity problem.
\end{remark}

\begin{remark}
\label{rem:avc_discont}
From the construction of the example above it follows immediately that the unassisted capacity of the classical AVC (without secrecy requirements) is discontinuous as well. To the best of our knowledge this has not been observed so far.
\end{remark}

An interesting observation appears when we analyze the previous result in the context of the different jamming strategies of the adversary, cf. Remark \ref{rem:strategies}. Obviously, the adversary is not able to choose a state sequence that will provide any information leakage to him, i.e., that security criterion cannot be attacked. Accordingly, the strategy will be to choose the state sequence in such a way that the legitimate communication is disturbed as much as possible. 

Now for the case $\lambda=0$ in Theorem \ref{the:avc_discont} we see the following: Whenever the legitimate users try to communicate at a positive rate, the adversary can jam the communication such that the decoding error at the legitimate receiver is always greater than $1/4$ since the $\avc$ is symmetrizable (see also \cite{BocheSchaefer13Superactivation} for more detailed discussion). Thus, no reliable communication is possible.

On the other hand, for $\lambda>0$ we have $C_S(\avwc(\lambda))>0$ so that in this case reliable and secure communication is possible. However, from this we cannot conclude on $C_S(\avwc(0))$ by taking the limit $\lambda\rightarrow0$, since $\lambda=0$ is a discontinuous point. Thus, it is not robust since small variations can result in a dramatic loss in secrecy capacity.

\subsection{Set of Non-Symmetrizable AVCs}
\label{sec:avc_continuity_region}

Here we show that it is possible to have non-trivial sets of AVWCs whose unassisted secrecy capacities are non-zero. 

For this purpose, we consider channels $\sigma,\sigma':\sX\rightarrow\sP(\sS)$, and an AVC given by the uncertainty set $\avc$. We define the function
\begin{equation*}
	F(\sigma,\sigma',\avc) = \sum_{x_1\in\sX}\sum_{x_2\in\sX}\sum_{y\in\sY}\big|W_\sigma(y|x_1,x_2)-W_{\sigma'}(y|x_1,x_2)\big|
\end{equation*}
with $W_\sigma(y|x_1,x_2)=\sum_{s\in\sS}W(y|x_1,s)\sigma(s|x_2)$. Then $F$ is a continuous function of $\sigma$, $\sigma'$, and $\avc$. 

\begin{lemma}
\label{lem:avc_lem1}
We have
\begin{equation*}
	\min_{\sigma,\sigma'}F(\sigma,\sigma',\avc) = 0
\end{equation*}
if and only if the AVC $\avc$ is symmetrizable.
\end{lemma}
\begin{IEEEproof}
If the AVC $\avc$ with the channel $\sigma:\sX\rightarrow\sP(\sS)$ is symmetrizable, then it obviously holds that $F(\sigma,\sigma',\avc)=0$. Thus, we only have to show the other direction.

Let $\hat{\sigma}$ and $\hat{\sigma}'$ be channels such that $F(\hat{\sigma},\hat{\sigma}',\avc)=0$. Then it also holds that $W_{\hat{\sigma}}(y|x_1,x_2) = W_{\hat{\sigma}'}(y|x_2,x_1)$ for all $y\in\sY$ and $x_1,x_2\in\sX$. We can interchange the elements $x_2$ and $x_1$ to obtain $W_{\hat{\sigma}}(y|x_2,x_1) = W_{\hat{\sigma}'}(y|x_1,x_2)$ for all $y\in\sY$ and $x_1,x_2\in\sX$ as well. Now, with
\begin{equation*}
	\tilde{\sigma}(s|x) = \frac{1}{2}\big(\hat{\sigma}(s|x)+\hat{\sigma}'(s|x)\big)
\end{equation*}
we obtain
\begin{align*}
	&\sum_{s\in\sS}W(y|x_2,s)\tilde{\sigma}(s|x_1) \\
	&\quad= \frac{1}{2}\Big(\sum_{s\in\sS}W(y|x_2,s)\hat{\sigma}(s|x_1) + \sum_{s\in\sS}W(y|x_2,s)\hat{\sigma}'(s|x_1)\Big) \\
	&\quad= \frac{1}{2}\big(W_{\hat{\sigma}}(y|x_2,x_1) + W_{\hat{\sigma}'}(y|x_2,x_1)\big) \\
	&\quad= \frac{1}{2}\big(W_{\hat{\sigma}'}(y|x_2,x_1) + W_{\hat{\sigma}}(y|x_1,x_2)\big) \\
	&\quad= \sum_{s\in\sS}\Big(\frac{1}{2}\big(\hat{\sigma}'(s|x_2) + \hat{\sigma}(s|x_2)\big)W(y|x_1,s)\Big) \\
	&\quad= \sum_{s\in\sS}W(y|x_1,s)\tilde{\sigma}(s|x_2)
\end{align*}
which shows that the AVC $\avc$ is symmetrizable, proving the lemma.
\end{IEEEproof}

\begin{lemma}
\label{lem:avc_lem2}
Let $\avc_1$ be an AVC with finite state set $\sS_1$ and
\begin{equation*}
	\min_{\sigma,\sigma'}F(\sigma,\sigma',\avc_1) = \delta > 0.
\end{equation*}
Then there exists an $\epsilon>0$ such that for all AVCs $\hat{\avc}_2$ with finite state set $\sS_2$ and $D(\avc_1,\avc_2)<\epsilon$ it holds that
\begin{equation*}
	\min_{\sigma,\sigma'}F(\sigma,\sigma',\hat{\avc}_2) \geq \frac{\delta}{2},
\end{equation*}
which means that these AVCs are also non-symmetrizable.
\end{lemma}
\begin{IEEEproof}
For every $s_2\in\sS_2$ there exists an $\hat{s}_1=\hat{s}_1(s_2)$ such that
\begin{equation*}
	\max_{x\in\sX}\sum_{y\in\sY}\big|\hat{W}(y|x,s_2) - W(y|x,\hat{s}_1)\big| < \epsilon.
\end{equation*}
Then it holds that
\begin{align*}
	&\Big|\sum_{s_2\in\sS_2}\hat{W}(y|x,s_2)\sigma(s_2|\hat{x}) - \sum_{s_2\in\sS_2}W(y|x,\hat{s}_1(s_2))\sigma(s_2|\hat{x})\Big| \\
	&\qquad\leq \sum_{s_2\in\sS_2}\big|\hat{W}(y|x,s_2) - W(y|x,\hat{s}_1(s_2))\big|\sigma(s_2|\hat{x}) \\
	&\qquad\leq \epsilon\sum_{s_2\in\sS_2}\sigma(s_2|\hat{x}) = \epsilon.
\end{align*}
Accordingly, we get
\begin{align}
	&\Big|\sum_{s_2\in\sS_2}\sigma(s_2|\hat{x})W(y|x,\hat{s}_1(s_2)) - \sum_{s_2\in\sS_2}\sigma'(s_2|x)W(y|\hat{x},\hat{s}_1(s_2))\Big| \nonumber \\
	&\quad= \Big|\sum_{s_2\in\sS_2}\sigma(s_2|\hat{x})\big(W(y|x,\hat{s}_1(s_2)) - \hat{W}(y|x,s_2)\big) \nonumber \\
	&\quad\quad - \sum_{s_2\in\sS_2}\sigma'(s_2|x)\big(W(y|\hat{x},\hat{s}_1(s_2)) - \hat{W}(y|\hat{x},s_2)\big) \nonumber \\
	&\quad\quad + \sum_{s_2\in\sS_2}\sigma(s_2|\hat{x})\hat{W}(y|\hat{x},s_2) - \sum_{s_2\in\sS_2}\sigma'(s_2|x)\hat{W}(y|\hat{x},s_2)\Big| \nonumber \\
	&\quad\leq 2\epsilon \!+\! \Big|\!\!\sum_{s_2\in\sS_2}\sigma(s_2|\hat{x})\hat{W}(y|x,s_2) -\!\! \sum_{s_2\in\sS_2}\sigma'(s_2|\hat{x})\hat{W}(y|x,s_2)\Big|.
	\label{eq:avc_lem2_distance}
\end{align}
Now, we define the channels $\tilde{\sigma},\tilde{\sigma}':\sX\rightarrow\sP(\sS_1)$ as follows: For fixed $s_1\in\sS_1$ let $\sI(s_1)$ be the set of all $s_2$ with $\hat{s}_1(s_2)=s_1$. Then we set
\begin{equation*}
	\tilde{\sigma}(s_1|x) = \sum_{s_2\in\sI(s_1)}\sigma(s_2|x).
\end{equation*}
Similarly, we define the channel $\tilde{\sigma}'$. Then left hand side of \eqref{eq:avc_lem2_distance} is equal to
\begin{equation*}
	\Big|\sum_{s_1\in\sS_1}\tilde{\sigma}(s_1|\hat{x})W(y|x,s_1) - \sum_{s_1\in\sS_1}\tilde{\sigma}'(s_1|x)W(y|\hat{x},s_1) \Big|
\end{equation*}
and therewith greater than $\delta$ per definition. Thus, we get
\begin{align}
	&\Big|\sum_{s_2\in\sS_2}\sigma(s_2|\hat{x})\hat{W}(y|x,s_2) - \sum_{s_2\in\sS_2}\sigma'(s_2|x)\hat{W}(y|\hat{x},s_2) \Big| \nonumber \\
	&\qquad\qquad\qquad\qquad\qquad\qquad\qquad\qquad\geq \delta - 2\epsilon.
	\label{eq:avc_lem2_distance2}
\end{align}
Finally, we set $\epsilon<\frac{\delta}{4}$ to complete the proof, since the right hand side of \eqref{eq:avc_lem2_distance2} is independent of $\sigma$ and $\sigma'$ so that we can equivalently choose the minimum.
\end{IEEEproof}
\vspace*{0.5\baselineskip}

With these lemmas we are in the position to prove the following result which shows that there are sets of AVWCs that all have a non-zero unassisted secrecy capacities.

\begin{theorem}
\label{the:avc_crancont}
Let $\avwc$ be an AVWC with finite state set such that $\Cavc>0$. Then there exists an $\epsilon>0$ such that for all AVWCs $\hat{\avwc}$ with finite state sets and $D(\avwc,\hat{\avwc})<\epsilon$ we have $C_S(\hat{\avwc})>0$.
\end{theorem}
\begin{IEEEproof}
Since $\Cavc>0$, we know from Theorem \ref{the:avc_detcapacity} that the corresponding AVC $\avc$ to the legitimate receiver is non-symmetrizable. Then from Lemma \ref{lem:avc_lem1} it follows that there must exist a $\delta>0$ such that $F(\sigma,\sigma',\avc)\geq\delta>0$. Since $\Cavccr$ is continuous, cf. Remark \ref{rem:continuity}, and $\Cavccr>0$, there exists an $\hat{\epsilon}>0$ such that for all AVWCs $\hat{\avwc}$ with finite state sets and $D(\avwc,\hat{\avwc})<\hat{\epsilon}$ we also have $C_{S,\text{CR}}(\hat{\avwc})>0$. Now, we choose an $\epsilon<\hat{\epsilon}$ in such a way that Lemma \ref{lem:avc_lem2} still holds. This implies that all AVCs $\hat{\avc}$ with finite state sets and $D(\avc,\hat{\avc})<\epsilon$ are non-symmetrizable so that we have $C_S(\hat{\avwc})=C_{S,\text{CR}}(\hat{\avwc})>0$.
\end{IEEEproof}
\vspace*{0.5\baselineskip}

\subsection{Set of Symmetrizable AVCs}
\label{sec:avc_continuity_sym}

In Section \ref{sec:avc_continuity_point} it was shown that the unassisted secrecy capacity of an AVWC can have a discontinuity point. Thus, symmetrizable AVCs in the legitimate link can appear as a discontinuity point within sets of non-symmetrizable AVCs. Here, we want to show that it is also possible that there are non-trivial sets of symmetrizable AVCs. 

To do so, we construct an example of an AVWC $\avwc^*$ such that, for a set of AVWCs $\avwc$ around this channel $\avwc^*$, we always have $\Cavccr>0$ and $\Cavc=C_S(\avwc^*)=0$. Therefore, we define the AVC to the legitimate receiver by the uncertainty set $\avc^*=\{W_1^*,W_2^*\}$ with
\begin{equation*}
	W_1^*=\begin{pmatrix}\frac{1}{2} & \frac{1}{2} & 0 \\ \frac{1}{4} & 0 & \frac{3}{4}\end{pmatrix} \quad\text{and}\quad
	W_2^*=\begin{pmatrix}0 & 0 & 1 \\ 0 & 1 & 0 \end{pmatrix}.
\end{equation*}
For the AVC to the eavesdropper we choose again $\avcv=\{V,V\}$ with $V$ as in \eqref{eq:avc_v}. With this choice, the channel to the eavesdropper remains fixed, while for the channel to the legitimate receiver we allow appropriate variations. The corresponding AVWC $\avwc^*$ is then $\avwc^*=\{\avc^*,\avcv\}$.

Further, we define the channel $\sigma^*:\sX\rightarrow\sP(\sS)$ as
\begin{equation*}
	\sigma^*=\begin{pmatrix}\frac{4}{5} & \frac{1}{5} \\ \frac{2}{5} & \frac{3}{5} \end{pmatrix}.
\end{equation*}
Then a simple calculation shows immediately that 
\begin{equation*}
	\sum_{s\in\sS}W^*_s(y|1)\sigma^*(s|2) = \sum_{s\in\sS}W^*_s(y|2)\sigma^*(s|1)
\end{equation*}
holds for all $y\in\sY$, which means that the AVC $\avc^*$ is symmetrizable so that $C_S(\avwc^*)=0$ by Theorem \ref{the:avc_detcapacity}. Now, the following theorem shows that there exists a set around this AVWC which has zero secrecy capacity as well.

\begin{theorem}
\label{the:avc_zeroregion}
There exists an $\epsilon_0>0$ such that for all AVWCs $\avwc=\{\avc,\avcv\}$ with finite state sets and $D(\avc^*,\avc)<\epsilon_0$ it always holds that
\begin{equation*}
	\Cavccr >0 \quad\text{and}\quad \Cavc = 0.
\end{equation*}
\end{theorem}
\begin{IEEEproof}
Due to Theorem \ref{the:avc_discont} it suffices to concentrate on the AVC $\avc$ to the legitimate receiver. Due to the choice of $V$, cf. \eqref{eq:avc_v}, for all AVWCs $\avwc$ we have 
\begin{equation}
	\Cavc = C(\avc)
	\label{eq:avc_zeroregion_1}
\end{equation}
and
\begin{equation}
	\Cavccr = C_{\text{CR}}(\avc).
	\label{eq:avc_zeroregion_2}
\end{equation}
Similarly as in the proof of Theorem \ref{the:avc_discont}, we can show that $C_{\text{CR}}(\avc^*)>0$. Since $C_{\text{CR}}(\cdot)$ is a continuous function, there exists an $\epsilon_0>0$ such that for all AVCs $\avc$ with finite state sets and $D(\avc^*,\avc)<\epsilon_0$ we have $C_{\text{CR}}(\avc)>0$ as well.

Due to relation \eqref{eq:avc_zeroregion_1} it remains to show that there exists an $\epsilon_1$ with $0\leq\epsilon_1\leq\epsilon_0$ such that all AVCs $\avc$ with finite state sets and $D(\avc^*,\avc)<\epsilon_1$ are symmetrizable. This can be shown similarly as in the proof of Theorem \ref{the:avc_discont} which then concludes the proof.
\end{IEEEproof}

\section{Robustness of Weak Secrecy Codes}
\label{sec:robust}

For the AVWC we have seen that the CR-assisted secrecy capacity $\Cavccr$ is continuous in the uncertainty set $\avwc$. In contrast to this, the unassisted secrecy capacity $\Cavc$ displays a discontinuous behavior. In particular, the previous discussion reveals that $\Cavc$ is always a continuous function of the eavesdropper channel, while the discontinuity comes from the legitimate channel only; see \cite{NoetzelXXAVWC} for further details. 

The fact that the secrecy capacity depends in a continuous way on the eavesdropper channel is a desirable property as it shows the robustness of the secrecy requirement. However, we cannot conclude from a continuous dependency of the secrecy capacity on the eavesdropper channel that any particular code itself displays this continuity as well. However, in the following we want to show that this is indeed the case.

We consider two AVWCs $\avwc_1=\{\avc,\avcv_1\}$ and $\avwc_2=\{\avc,\avcv_2\}$ which share the same AVC $\avc$ to the legitimate receiver, but consist of different AVCs $\avcv_1$ and $\avcv_2$ to the eavesdropper. The question we want to explore is now the following: Is a ``good'' code (which realizes secrecy) for $\avwc_1$ also a ``good'' code for $\avwc_2$ when both eavesdropper channels are close (i.e., $D(\avcv_1,\avcv_2)\leq\epsilon$ for some small $\epsilon>0$)? We will answer this for the weak secrecy criterion for which \eqref{eq:avc_achievable2} is replaced by
\begin{equation*}
	\max_{s^n\in\sS^n}\frac{1}{n}I(J;Z_{s^n}^n\|\Cdet_n) \leq \delta_n.
\end{equation*}
For this purpose, we need the following lemma.
\begin{lemma}
\label{lem:avc_lem3}
For $n\in\N$ arbitrary let $V_m,\widetilde{V}_m:\sX\rightarrow\sP(\sZ)$, $m\in\{1,...,n\}$, be channels with 
\begin{equation*}
	d(V_m,\widetilde{V}_m)\leq\epsilon
\end{equation*}
for some $\epsilon>0$. Let $\sU$ be an arbitrary finite set, $P_U\in\sP(\sU)$ the uniform distribution on $\sU$, and $E(x^n|u)$, $x^n\in\sX^n$, an arbitrary stochastic encoder, cf. \eqref{eq:cc_encoder}. We consider probability distributions
\begin{align*}
	P_{UZ^n} &= \sum_{x^n\in\sX^n}\prod_{m=1}^nV_m(z_m|x_m)E(x^n|u)P_U(u) \\
	\Ptil_{UZ^n} &= \sum_{x^n\in\sX^n}\prod_{m=1}^n\widetilde{V}_m(z_m|x_m)E(x^n|u)P_U(u).
\end{align*}
Then it holds that
\begin{equation}
	\big|I(U;Z^n\|P) - I(U;Z^n\|\Ptil)\big| \leq n\delta_2(\epsilon,|\sZ|) 
\end{equation}
with $\delta_2(\epsilon,|\sZ|)$ as in Lemma \ref{lem:cc_lem2}, cf. \eqref{lem:cc_lem2}.
\end{lemma}
\begin{IEEEproof}
The proof follows by a suitable adaptation of the proof of Lemma \ref{lem:cc_lem2}. This is sketched in Appendix \ref{sec:app_proof_lem3}.
\end{IEEEproof}
\vspace*{0.5\baselineskip}

Now we are in the position to show that a code that realizes secrecy over a certain AVC, is also a ``good'' code for all AVCs in a certain neighborhood.

\begin{theorem}
\label{the:avc_robustness}
Let $\avcv$ with finite state set $\sS$ be an AVC to the eavesdropper and for $n\in\N$ let $\Cdet_n$ be an unassisted code that achieves weak secrecy
\begin{equation}
	\max_{s^n\in\sS^n}\frac{1}{n}I(J;Z_{s^n}^n\|\Cdet_n) = \delta_n.
	\label{eq:avc_leakage1}
\end{equation}
Then it holds for all AVCs $\avcv^*$ with finite state sets $\sS^*$ and $D(\avcv,\avcv^*)\leq\epsilon$ that
\begin{equation}
	\max_{s_*^n\in\sS_*^n}\frac{1}{n}I(J;Z_{s^n_*}^n\|\Cdet_n) < \delta_n + \delta_2(\epsilon,|\sZ|)
	\label{eq:avc_leakage2}
\end{equation}
with $\delta_2(\epsilon,|\sZ|)$ as in Lemma \ref{lem:cc_lem2}, cf. \eqref{eq:cc_lem2}.
\end{theorem}
\begin{IEEEproof}
Let $\avcv^*$ with finite state set $\sS^*$ be an arbitrary AVC that satisfies $D(\avcv,\avcv^*)<\epsilon$. Further, let $s_*^n=(s_{1,*},s_{2,*},...,s_{n,*})\in\sS^*$ be arbitrary. Then for every $s_{m,*}$, $m\in\{1,...,n\}$, there exists an $s_m=s_m(s_{m,*})\in\sS$ such that
\begin{equation*}
	d(V_{s_{m,*}},V_{s_m})<\epsilon.
\end{equation*}
This allows us to apply Lemma \ref{lem:avc_lem3} from which we obtain
\begin{equation*}
	\big|\frac{1}{n}I(J;Z_{s^n}^n\|\Cdet_n)-\frac{1}{n}I(J;Z_{s^n(s_*^n)}^n\|\Cdet_n)\big| < \delta_2(\epsilon,|\sZ|)
\end{equation*}
so that
\begin{align*}
	\frac{1}{n}I(J;Z_{s_*^n}^n\|\Cdet_n) 
		&\leq \big|\frac{1}{n}I(J;Z_{s_*^n}^n\|\Cdet_n)-\frac{1}{n}I(J;Z_{s^n(s_*^n)}^n\|\Cdet_n)\big| \\
		&\qquad\qquad+ \frac{1}{n}I(J;Z_{s^n(s_*^n)}^n\|\Cdet_n) \\
		&< \delta_n + \delta_2(\epsilon,|\sZ|)
\end{align*}
which proves the result.
\end{IEEEproof}
\vspace*{0.5\baselineskip}

Next, we want to establish a similar result for CR-assisted codes as well.

\begin{theorem}
\label{the:avc_robustnesscr}
Let $\avcv$ with finite state set $\sS$ be an AVC to the eavesdropper and for $n\in\N$ let $\sC_{\text{CR},n}$ be a CR-assisted code that achieves weak secrecy
\begin{equation}
	\max_{s^n\in\sS^n}\frac{1}{n}I(J;Z_{s^n}^n\|\sC_{\text{CR},n}) = \delta_n.
	\label{eq:avc_leakage3}
\end{equation}
Then it holds for all AVCs $\avcv^*$ with finite state sets $\sS^*$ and $D(\avcv,\avcv^*)\leq\epsilon$ that
\begin{equation}
	\max_{s_*^n\in\sS_*^n}\frac{1}{n}I(J;Z_{s^n_*}^n\|\sC_{\text{CR},n}) < \delta_n + \delta_2(\epsilon,|\sZ|)
	\label{eq:avc_leakage4}
\end{equation}
with $\delta_2(\epsilon,|\sZ|)$ as in Lemma \ref{lem:cc_lem2}, cf. \eqref{eq:cc_lem2}.
\end{theorem}
\begin{IEEEproof}
As in Theorem \ref{the:avc_robustness} let  $s_*^n=(s_{1,*},s_{2,*},...,s_{n,*})\in\sS^*$ be arbitrary. Then for every $s_{m,*}$, $m\in\{1,...,n\}$, there exists an $s_m=s_m(s_{m,*})\in\sS$ such that $d(V_{s_{m,*}},V_{s_m})<\epsilon$. Then Lemma \ref{lem:avc_lem3} immediately yields $|\frac{1}{n}I(J;Z_{s^n}^n\|\sC_{\text{CR},n})-\frac{1}{n}I(J;Z_{s^n(s_*^n)}^n\|\sC_{\text{CR},n})| < \delta_2(\epsilon,|\sZ|)$ so that
\begin{align*}
	&\frac{1}{n}I(J;Z_{s_*^n}^n\|\sC_{\text{CR},n}) = \frac{1}{n}\sum_{\gamma\in\sG_n}I(J;Z_{s_*^n}^n\|\sC_n(\gamma))P_\Gamma(\gamma)\\
		&\;\leq \Big|\frac{1}{n}\sum_{\gamma\in\sG_n}\!\!P_\Gamma(\gamma)\big(I(J;Z_{s_*^n}^n\|\sC_n(\gamma))-I(J;Z_{s^n(s_*^n)}^n\|\sC_n(\gamma))\big)\Big| \\
		&\qquad\qquad+ \frac{1}{n}\sum_{\gamma\in\sG_n}I(J;Z_{s^n(s_*^n)}^n\|\sC_n(\gamma))P_\Gamma(\gamma) \\
		&\;\leq \frac{1}{n}\sum_{\gamma\in\sG_n}\!\!P_\Gamma(\gamma)\Big|I(J;Z_{s_*^n}^n\|\sC_n(\gamma))-I(J;Z_{s^n(s_*^n)}^n\|\sC_n(\gamma))\Big| + \delta_n \\
		&\;\leq \sum_{\gamma\in\sG_n}\!\!P_\Gamma(\gamma)\delta_n + \delta_2(\epsilon,|\sZ|) < \delta_n + \delta_2(\epsilon,|\sZ|) 
\end{align*}
which proves the result.
\end{IEEEproof}
\vspace*{0.5\baselineskip}

These results show that unassisted and CR-assisted codes are indeed robust in the weak secrecy sense. A code with a small information leakage rate for $\avcv$ as in \eqref{eq:avc_leakage1} or \eqref{eq:avc_leakage3} has also a small information leakage for all AVCs $\avcv^*$ with $D(\avcv,\avcv^*)\leq\epsilon$ as in \eqref{eq:avc_leakage2} or \eqref{eq:avc_leakage4}, respectively. In addition, the change in information leakage is explicitly quantified by $\delta_2(\epsilon,|\sZ|)$ in \eqref{eq:avc_leakage2} and \eqref{eq:avc_leakage4}. Note that, the robustness of codes immediately imply that the secrecy capacity is continuous in the eavesdropper channel as well.

\section{Conclusion}
\label{sec:conclusion}

In this paper we have considered secure communication over compound and arbitrarily varying channels. The analysis of this paper was motivated by the question of whether the secrecy capacity depends continuously on the uncertainty set or not. Obviously, a continuous behavior is desirable as then small changes in the uncertainty set result in only small changes in the secrecy capacity. This becomes particularly relevant in the context of active adversaries where the uncertainty set describes the strategy space of an adversary.

Surprisingly, the answer to the question of whether the secrecy is continuous or not depends highly on the abilities of the adversary -- even for the simplest case of an uncertainty set containing two elements. If the actual realization from this uncertainty set remains constant for the whole duration of the transmission, the scenario at hand is the compound wiretap channel and the corresponding secrecy capacity is a continuous function of this uncertainty set. However, if the adversary is allowed to vary in an unknown and arbitrary manner between these two realizations during the transmission, the legitimate users have to deal with an AVWC and its unassisted secrecy capacity can be discontinuous in the uncertainty set. More sophisticated strategies based on common randomness can help to overcome this discontinuity problem and the corresponding CR-assisted secrecy capacity of the AVWC is a continuous function of the uncertainty set.

\appendix

\subsection{Proof of Lemma \ref{lem:cc_lem1}}
\label{sec:app_proof_lem1}

The proof of the lemma follows \cite{Alicki04ContinuityQuantumConditionalInformation} where a similar result is presented in the context of quantum information. However, in contrast to the quantum version in \cite{Alicki04ContinuityQuantumConditionalInformation} we are able to get a better constant by using the fact that $H(Y|X\|P_{XY})\geq0$ for all $P_{XY}\in\sP(\sX\times\sY)$.

Let $P_{XY},Q_{XY}\in\sP(\sX\times\sY)$ with $\|P_{XY}-Q_{XY}\|\leq\epsilon$. We assume that
\begin{equation}
	\sum_{x\in\sX}\sum_{y\in\sY}|P_{XY}(x,y)-Q_{XY}(x,y)|=\epsilon
	\label{eq:cc_lem1_epsilon}
\end{equation}
is satisfied with equality since otherwise we could replace $\epsilon$ in \eqref{eq:cc_lem1_epsilon} with a smaller $\tilde{\epsilon}<\epsilon$ accordingly. 

We define
\begin{equation}
	f(x,y) \coloneqq |P_{XY}(x,y)-Q_{XY}(x,y)|
	\label{eq:cc_lem1_f}
\end{equation}
and set
\begin{equation*}
	p^*(x,y) = (1-\epsilon)P_{XY}(x,y)+f(x,y)
\end{equation*}
for all $(x,y)\in\sX\times\sY$ so that $p^*\in\sP(\sX,\sY)$ is a joint probability distribution on $\sX\times\sY$.

Further, we set
\begin{subequations}
\label{eq:cc_lem1_pqhat}
\begin{align}
	\hat{p}(x,y) &= \frac{1}{\epsilon}f(x,y), \\
	\hat{q}(x,y) &= \frac{1}{\epsilon}\big((1-\epsilon)\big[P_{XY}(x,y)-Q_{XY}(x,y)\big]+f(x,y)\big).
\end{align}
\end{subequations}
We have to check that $\hat{p}$ and $\hat{q}$ are well defined in the sense that they are probability distributions. $\hat{p}(x,y)\geq0$ for all $(x,y)\in\sX\times\sY$ is obviously true. It remains to check that $\hat{q}(x,y)\geq0$ for all $(x,y)\in\sX\times\sY$ is also satisfied.

If $P_{XY}(x,y)\leq Q_{XY}(x,y)$, then
\begin{align*}
	-f(x,y) &\leq P_{XY}(x,y)-Q_{XY}(x,y) \\
		&\leq (1-\epsilon)\big(P_{XY}(x,y)-Q_{XY}(x,y)\big) \leq 0
\end{align*}
so that $\hat{q}(x,y)\geq0$. On the other hand, if $P_{XY}(x,y)>Q_{XY}(x,y)$, then
\begin{align*}
	0 &<(1-\epsilon)\big(P_{XY}(x,y)-Q_{XY}(x,y)\big) \\
		&\leq P_{XY}(x,y)-Q_{XY}(x,y) \leq f(x,y)
\end{align*}
so that $\hat{q}(x,y)\geq0$ also in this case. Further, from the definition of $\hat{p}$ and $\hat{q}$ in \eqref{eq:cc_lem1_pqhat} and \eqref{eq:cc_lem1_epsilon}-\eqref{eq:cc_lem1_f} it can easily be verified that
\begin{equation*}
	\sum_{x\in\sX}\sum_{y\in\sY}\hat{p}(x,y)= \sum_{x\in\sX}\sum_{y\in\sY}\hat{q}(x,y)=1
\end{equation*}
so that $\hat{p}\in\sP(\sX\times\sY)$ and $\hat{q}\in\sP(\sX\times\sY)$ are joint probability distributions.

With this we can express $p^*$ as
\begin{subequations}
\begin{align}
	p^*(x,y) &= (1-\epsilon)P_{XY}(x,y) + \epsilon\hat{p}(x,y) \label{eq:cc_lem1_pstar1}\\
		&= (1-\epsilon)Q_{XY}(x,y) + \epsilon\hat{q}(x,y) \label{eq:cc_lem1_pstar2}
\end{align}
\end{subequations}
for all $(x,y)\in\sX\times\sY$. Next, we show that \eqref{eq:cc_lem1_pstar1} implies
\begin{equation}
	\big|H(Y|X\|P_{XY})-H(Y|X\|p^*)\big| \leq \epsilon\log|\sY|+H_2(\epsilon).
	\label{eq:cc_lem1_pstar3}
\end{equation}
To do so, we use the fact that the conditioned entropy is concave, i.e.,
\begin{equation*}
	H(Y|X\|p^*) \geq (1-\epsilon)H(Y|X\|P_{XY}) + \epsilon H(Y|X\|\hat{p}).
\end{equation*}
With this, we have
\begin{align}
	&H(Y|X\|P_{XY}) - H(Y|X\|p^*) \nonumber \\
	&\quad\leq H(Y|X\|P_{XY}) - (1-\epsilon)H(Y|X\|P_{XY}) - \epsilon H(Y|X\|\hat{p}) \nonumber \\
	&\quad= \epsilon\big(H(Y|X\|P_{XY})-H(Y|X\|\hat{p})\big) \nonumber \\
	&\quad\leq \epsilon H(Y|X\|P_{XY}) \leq \epsilon\log|\sY|.
	\label{eq:cc_lem1_hdiff1}
\end{align}
Using the concavity of the entropy
\begin{equation*}
	H(X\|p^*) \geq (1-\epsilon)H(X\|P_{XY}) + \epsilon H(X\|\hat{p})
\end{equation*}
and the upper bound on the joint entropy
\begin{equation*}
	H(X,Y\|p^*) \leq (1-\epsilon)H(X,Y\|P_{XY})+\epsilon H(X,Y\|\hat{p})+ H_2(\epsilon),
\end{equation*}
we get
\begin{align*}
	H(Y|X\|p^*) &= H(X,Y\|p^*) - H(X\|p^*) \\
	&\leq (1-\epsilon)H(Y|X\|P_{XY}) + \epsilon H(Y|X\|p^*) + H_2(\epsilon)
\end{align*}
and further
\begin{align}
	&H(Y|X\|P_{XY}) - H(Y|X\|p^*) \nonumber \\
	&\quad\geq -\epsilon\big(H(Y|X\|p^*) - H(Y|X\|P_{XY})\big) - H_2(\epsilon) \nonumber \\
	&\quad\geq -\epsilon H(Y|X\|p^*) - H_2(\epsilon) \geq -\epsilon\log|\sY| - H_2(\epsilon).
	\label{eq:cc_lem1_hdiff2}
\end{align}
Now, \eqref{eq:cc_lem1_hdiff1} and \eqref{eq:cc_lem1_hdiff2} yield
\begin{equation*}
	\big|H(Y|X\|P_{XY}) - H(Y|X\|p^*)\big| \leq \epsilon\log|\sY| + H_2(\epsilon)
\end{equation*}
which shows \eqref{eq:cc_lem1_pstar3}. (By the same arguments, one can show that \eqref{eq:cc_lem1_pstar2} implies $|H(Y|X\|Q_{XY})-H(Y|X\|p^*)| \leq \epsilon\log|\sY|+H_2(\epsilon)$.)

Finally, this yields
\begin{align*}
	&\big|H(Y|X\|P_{XY})-H(Y|X\|Q_{XY})\big| \\
	&\qquad= \big|H(Y|X\|P_{XY})-H(Y|X\|p^*) \\
	&\qquad\qquad+ \big(H(Y|X\|p^*)-H(Y|X\|Q_{XY})\big)\big| \\
	&\qquad\leq \big|H(Y|X\|P_{XY})-H(Y|X\|p^*)\big| \\
	&\qquad\qquad + \big|H(Y|X\|Q_{XY})-H(Y|X\|p^*)\big| \\
	&\qquad\leq 2\epsilon\log|\sY| + 2H_2(\epsilon)
\end{align*}
which is \eqref{eq:cc_lem1} proving the lemma. \hfill\IEEEQED

\subsection{Proof of Lemma \ref{lem:cc_lem2}}
\label{sec:app_proof_lem2}

Let $0\leq k\leq n$ arbitrary. We define
\begin{equation}
	P_{UY^n}^{(k)}(u,y^n) \coloneqq\!\!\! \sum_{x^n\in\sX^n}\prod_{l=1}^kW(y_l|x_l)\!\!\prod_{l=k+1}^n\!\!\!\Wtil(y_l|x_l)E(x^n|u)P_U(u).
	\label{eq:cc_lem2_puy}
\end{equation}
Then it holds that $P_{UY^n}^{(0)} = \Ptil_{UY^n}$ and$P_{UY^n}^{(n)} = P_{UY^n}$. Now we have
\begin{align}
	&I(U;Y^n\|P_{UY^n}^{(n)}) - I(U;Y^n\|P_{UY^n}^{(0)}) \nonumber \\
	&\qquad\qquad= \sum_{k=0}^{n-1}\big(I(U;Y^n\|P_{UY^n}^{(k+1)}-I(U;Y^n\|P_{UY^n}^{(k)})\big).
	\label{eq:cc_lem2_sum}
\end{align}
For all $0\leq k \leq n-1$ it holds that
\begin{align}
	&I(U;Y^n\|P_{UY^n}^{(k+1)})-I(U;Y^n\|P_{UY^n}^{(k)}) \nonumber\\
	&\qquad =H(Y^n\|P_{UY^n}^{(k+1)})-H(Y^n\|P_{UY^n}^{(k)}) \nonumber\\
	&\qquad\qquad- H(U,Y^n\|P_{UY^n}^{(k+1)})+H(U,Y^n\|P_{UY^n}^{(k)}).
	\label{eq:cc_lem2_hk}
\end{align}
We want to analyze the right hand side of \eqref{eq:cc_lem2_hk} in more detail. We present the analysis for the second expression, the other one follows by the same arguments. For $0\leq k\leq n-1$ we have
\begin{align}
	&\sum_{u\in\sU}\sum_{y^n\in\sY^n}\big|P_{UY^n}^{(k+1)}(u,y^n)-P_{UY^n}^{(k)}(u,y^n)\big| \nonumber \\
	&\quad= \sum_{u\in\sU}\sum_{y^n\in\sY^n}\bigg|\sum_{x^n\in\sX^n}\Big(\prod_{l=1}^{k+1}W(y_l|x_l)\prod_{l=k+2}^n\Wtil(y_l|x_l) \nonumber \\
	&\quad\quad- \prod_{l=1}^{k}W(y_l|x_l)\prod_{l=k+1}^n\Wtil(y_l|x_l)\Big)E(x^n|u)P_U(u)\bigg| \nonumber \\
	&\quad= \sum_{u\in\sU}\sum_{y^n\in\sY^n}\bigg|\sum_{x^n\in\sX^n}\prod_{l=1}^kW(y_l|x_l)\prod_{l=k+2}^n\Wtil(y_l|x_l) \nonumber \\
	&\quad\quad\times\Big(W(y_{k+1}|x_{k+1})-\Wtil(y_{k+1}|x_{k+1})\Big)E(x^n|u)P_U(u)\bigg| \nonumber \\
	&\quad\leq \sum_{u\in\sU}\sum_{y^n\in\sY^n}\sum_{x^n\in\sX^n}\prod_{l=1}^kW(y_l|x_l)\prod_{l=k+2}^n\Wtil(y_l|x_l) \nonumber \\
	&\quad\qquad\times\Big|W(y_{k+1}|x_{k+1})-\Wtil(y_{k+1}|x_{k+1})\Big| E(x^n|u)P_U(u) \nonumber \\
	&\quad= \sum_{u\in\sU}\sum_{x^n\in\sX^n}\bigg(\sum_{y^n\in\sY^n}\prod_{l=1}^kW(y_l|x_l)\prod_{l=k+2}^n\Wtil(y_l|x_l) \nonumber \\
	&\quad\quad\times\Big|W(y_{k+1}|x_{k+1})-\Wtil(y_{k+1}|x_{k+1})\Big|\bigg)E(x^n|u)P_U(u) \nonumber \\
	&\quad= \sum_{u\in\sU}\sum_{x^n\in\sX^n}\sum_{y_{k+1}\in\sY}\Big|W(y_{k+1}|x_{k+1})-\Wtil(y_{k+1}|x_{k+1})\Big| \nonumber \\
	&\quad\quad \times E(x^n|u)P_U(u) \nonumber \\
	&\quad< \epsilon\cdot\sum_{u\in\sU}\sum_{x^n\in\sX^n}E(x^n|u)P_U(u) = \epsilon
	\label{eq:cc_lem2_ppksmall}
\end{align}
where \eqref{eq:cc_lem2_ppksmall} follows from the fact that the distance between $W$ and $\Wtil$ is small by assumption, cf. \eqref{eq:cc_lem2_d}. Thus, \eqref{eq:cc_lem2_ppksmall} shows that the total variation between the joint probability distributions $P_{UY^n}^{(k)}$ and $P_{UY^n}^{(k+1)}$ is smaller than $\epsilon$.

We define $Y_{-k}^n \coloneqq (Y_1,Y_2,...,Y_{k-1},Y_{k+1},...,Y_n)$ as the sequence without the $k$-th element. In the following we study the random variables $(U,Y_{-(k+1)}^n)$ which are distributed according to $P_{UY^n}^{(k+1)}$, i.e., they have the probability distribution
\begin{align}
	&P_{UY_{-(k+1)}^n}^{(k+1)}(u,y_{-(k+1)}^n) = \sum_{y_{k+1}\in\sY}P_{UY^n}^{(k+1)}(u,y^n) \nonumber \\
	&\quad= \sum_{y_{k+1}\in\sY}\sum_{x^n\in\sX^n}\prod_{l=1}^kW(y_l|x_l)W(y_{k+1}|x_{k+1}) \nonumber \\
	&\quad\qquad \times\prod_{l=k+2}^n\Wtil(y_l|x_l)E(x^n|u)P_U(u) \nonumber \\
	&\quad= \sum_{x^n\in\sX^n}\prod_{l=1}^kW(y_l|x_l)\prod_{l=k+2}^n\Wtil(y_l|x_l)E(x^n|u)P_U(u).
	\label{eq:cc_lem2_puy2}
\end{align}
Similarly, we look at the random variables $(U,Y_{-(k+1)}^n)$ which are distributed according to $P_{UY^n}^{(k)}$, i.e.,
\begin{align}
	&P_{UY_{-(k+1)}^n}^{(k)}(u,y_{-(k+1)}^n) = \sum_{y_{k+1}\in\sY}P_{UY^n}^{(k)}(u,y^n) \nonumber \\
	&\quad= \sum_{y_{k+1}\in\sY}\sum_{x^n\in\sX^n}\prod_{l=1}^kW(y_l|x_l)\Wtil(y_{k+1}|x_{k+1}) \nonumber \\
	&\quad\qquad\times\prod_{l=k+2}^n\Wtil(y_l|x_l)E(x^n|u)P_U(u) \nonumber \\
	&\quad= \sum_{x^n\in\sX^n}\prod_{l=1}^kW(y_l|x_l)\prod_{l=k+2}^n\Wtil(y_l|x_l)E(x^n|u)P_U(u) \nonumber \\
	&\quad= P_{UY_{-(k+1)}^n}^{(k+1)}(u,y_{-(k+1)}^n)
	\label{eq:cc_lem2_puy3}
\end{align}
and observe that $P_{UY^n}^{(k)}$ and $P_{UY^n}^{(k+1)}$ have the same marginal distribution for $(U,Y_{-(k+1)}^n)$. This implies that
\begin{equation}
	H(U,Y_{-(k+1)}^n\|P_{UY^n}^{(k+1)}) = H(U,Y_{-(k+1)}^n\|P_{UY^n}^{(k)})
	\label{eq:cc_lem2_hhk}
\end{equation}
so that
\begin{align}
	&\big|H(U,Y^n\|P_{UY^n}^{(k+1)}) - H(U,Y^n\|P_{UY^n}^{(k)})\big| \nonumber \\
	&\;= \big|H(U,Y^n\|P_{UY^n}^{(k+1)}) - H(U,Y_{-(k+1)}^n\|P_{UY^n}^{(k+1)}) \nonumber \\
	&\;\qquad + H(U,Y_{-(k+1)}^n\|P_{UY^n}^{(k)}) - H(U,Y^n\|P_{UY^n}^{(k)})\big| \nonumber \\
	&\;= \big|H(Y_{k+1}|U,Y_{-(k+1)}^n\|P_{UY^n}^{(k+1)}) \nonumber\\
	&\;\qquad - H(Y_{k+1}|U,Y_{-(k+1)}^n\|P_{UY^n}^{(k)})\big| \nonumber \\
	&\;< 2\epsilon\log|\sY|+2H_2(\epsilon)
	\label{eq:cc_lem2_huy}
\end{align}
where the first step follows from \eqref{eq:cc_lem2_hhk} and the last step from Lemma \ref{lem:cc_lem1} and \eqref{eq:cc_lem2_ppksmall}. Similarly, using the same ideas one can easily show that 
\begin{equation}
	\big|H(Y^n\|P_{UY^n}^{(k+1)}) - H(Y^n\|P_{UY^n}^{(k)})\big| < 2\epsilon\log|\sY|+2H_2(\epsilon) 
	\label{eq:cc_lem2_hy}
\end{equation}
holds as well.

Now, inserting \eqref{eq:cc_lem2_huy} and \eqref{eq:cc_lem2_hy} into \eqref{eq:cc_lem2_hk}, we obtain $|I(U;Y^n\|P_{UY^n}^{(k+1)})-I(U;Y^n\|P_{UY^n}^{(k)})|\leq 4\epsilon\log|\sY|+4H_2(\epsilon)\eqqcolon\delta_2(\epsilon,|\sY|)$
so that the expression in \eqref{eq:cc_lem2_sum} becomes
\begin{align*}
	&\big|I(U;Y^n\|P_{UY^n}^{(n)})-I(U;Y^n\|P_{UY^n}^{(0)})\big| \\
	&\quad\leq \sum_{k=0}^{n-1}\big|I(U;Y^n\|P_{UY^n}^{(k+1)})-I(U;Y^n\|P_{UY^n}^{(k)})\big| \leq n\delta_2(\epsilon,|\sY|)
\end{align*}
which proves the lemma. \hfill\IEEEQED

\subsection{Proof of Lemma \ref{lem:avc_lem3}}
\label{sec:app_proof_lem3}

In the previous Lemma \ref{lem:cc_lem2} the channels remains the same for the whole block length $n$. In contrast to that, the current Lemma \ref{lem:avc_lem3} allows different channels for each time instant. 

Investigating the proof of Lemma \ref{lem:cc_lem2} reveals that the proof remains valid when we replace the channels $W$ and $\widetilde{W}$ by the new channels $V_m$ and $\widetilde{V}_m$, where $m\in\{1,...,n\}$ is chosen according to the corresponding time instant in \eqref{eq:cc_lem2_puy}, \eqref{eq:cc_lem2_ppksmall}, \eqref{eq:cc_lem2_puy2}, and \eqref{eq:cc_lem2_puy3}. Then the desired result follows immediately. \hfill\IEEEQED

\section*{Acknowledgment}

This work was motivated by discussions at the ``BSI-Workshop on Physical Layer Security'' at the Federal Office for Information Security (BSI), Bonn, Germany, Feb. 2013 and the industrial board meeting on ``Information Security'' of the German Ministry of Education and Research (BMBF) in Bonn, Germany, May 2013. H. Boche would like to thank Dr. R. Baumgart, secunet Security Networks Inc., and Dr. R. Plaga, BSI, for motivating and fruitful discussions that led to these results.


\begin{IEEEbiography}
[{\includegraphics[width=1in,height=1.25in,clip,keepaspectratio]{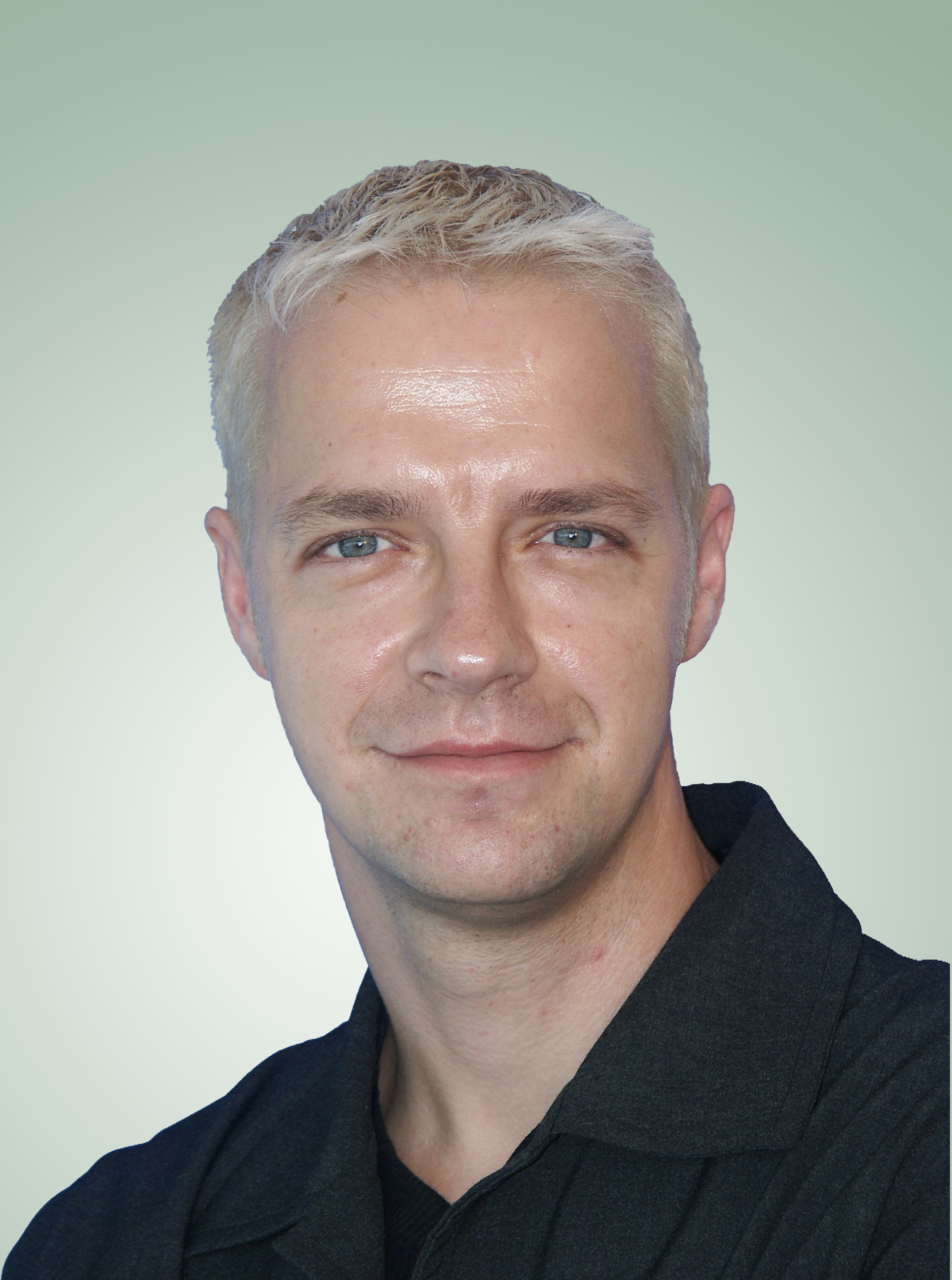}}]{Holger Boche}
(M'04--SM'07--F'11) received the Dr.rer.nat. degree in pure mathematics from the Technische Universit\"at Berlin, Berlin, Germany, in 1998, the Dipl.-Ing. and Dr.-Ing. degrees in electrical engineering from the Technische Universit\"at Dresden, Dresden, Germany, in 1990 and 1994, respectively, and the degree in mathematics from the Technische Universit\"at Dresden, in 1992. From 1994 to 1997, he was involved in post-graduate studies in mathematics with the Friedrich-Schiller Universit\"at Jena, Jena, Germany. In 1997, he joined the Heinrich-Hertz-Institut (HHI) fu\"ur Nachrichtentechnik Berlin, Berlin. In 2002, he was a Full Professor of Mobile Communication Networks with the Institute for Communications Systems, Technische Universit\"at Berlin. In 2003, he became the Director of the Fraunhofer German-Sino Laboratory for Mobile Communications, Berlin, and the Director of HHI in 2004. He was a Visiting Professor with ETH Zurich, Zurich, Switzerland, in Winter 2004 and 2006, and KTH Stockholm, Stockholm, Sweden, in Summer 2005. Since 2010, he has been with the Institute of Theoretical Information Technology and a Full Professor with the Technische Universit\"at M\"ünchen, Munich, Germany. Since 2014, he has been a member and an Honorary Fellow of the TUM Institute for Advanced Study, Munich. He is a member of the IEEE Signal Processing Society SPCOM and the SPTM Technical Committee. He received the Research Award Technische Kommunikation from the Alcatel SEL Foundation in 2003, the Innovation Award from the Vodafone Foundation in 2006, and the Gottfried Wilhelm Leibniz Prize from the German Research Foundation in 2008. He was a corecipient of the 2006 IEEE Signal Processing Society Best Paper Award and a recipient of the 2007 IEEE Signal Processing Society Best Paper Award. He was elected as a member of the German Academy of Sciences (Leopoldina) in 2008 and the Berlin Brandenburg Academy of Sciences and Humanities in 2009.
\end{IEEEbiography}

\begin{IEEEbiography}
[{\includegraphics[width=1in,height=1.25in,clip,keepaspectratio]{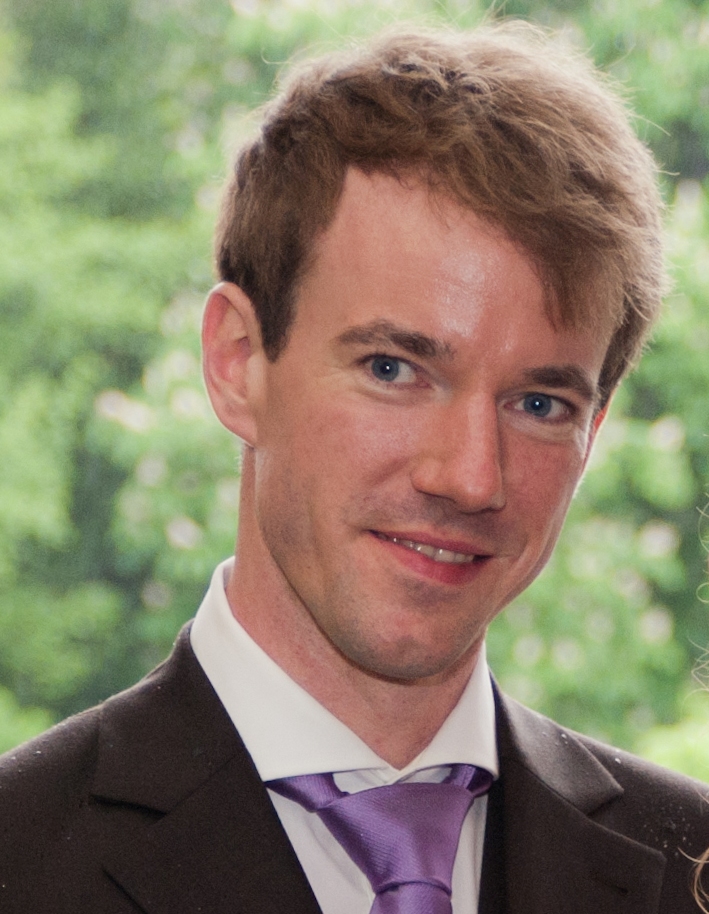}}]{Rafael F. Schaefer}
(S'08--M'12) received the Dipl.-Ing. degree in electrical engineering and computer science from the Technische Universit\"at Berlin, Berlin, Germany, in 2007, and the Dr.-Ing. degree in electrical engineering from the Technische Universit\"at M\"unchen, Munich, Germany, in 2012. He was a Research and Teaching Assistant with the Heinrich-Hertz-Lehrstuhl für Mobilkommunikation, Technische Universit\"at Berlin, from 2007 to 2010, and the Lehrstuhl f\"ur Theoretische Informationstechnik, Technische Universit\"at M\"unchen, from 2010 to 2013. He is currently a Post-Doctoral Research Fellow with the Department of Electrical Engineering, Princeton University, Princeton, NJ, USA. He was a recipient of the VDE Johann-Philipp-Reis Prize in 2013. He was one of the exemplary reviewers of the \textsc{IEEE Communication Letters} in 2013. He is currently an Associate Member of the IEEE Information Forensics and Security Technical Committee.
\end{IEEEbiography}

\begin{IEEEbiography}
[{\includegraphics[width=1in,height=1.25in,clip,keepaspectratio]{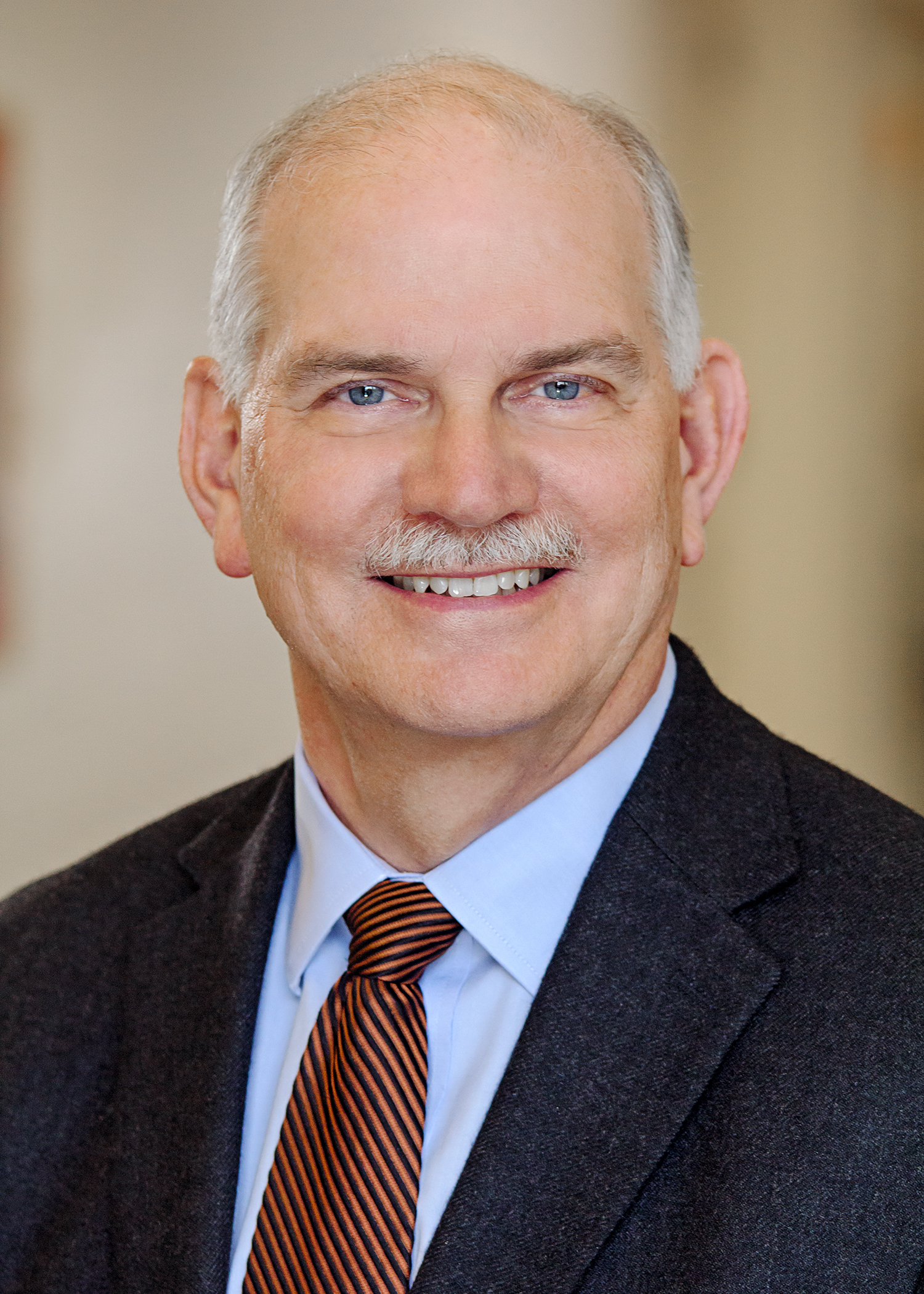}}]{H. Vincent Poor}
(S'72--M'77--SM'82--F'87) received the Ph.D. degree in electrical engineering and computer science from Princeton University, in 1977. From 1977 to 1990, he was on the faculty of the University of Illinois at Urbana–Champaign. Since 1990, he has been a Faculty Member with Princeton University, where he is currently the Dean of Engineering and Applied Science, and the Michael Henry Strater University Professor of Electrical Engineering. He has also held visiting appointments at several other institutions, most recently with Imperial College and Stanford University. His research interests are in the areas of information theory, stochastic analysis and statistical signal processing, and their applications in wireless networks and related fields. Among his publications in these areas is the recent book \textit{Mechanisms and Games for Dynamic Spectrum Allocation} (Cambridge University Press, 2014).

Dr. Poor is a member of the National Academy of Engineering and the National Academy of Sciences, and a Foreign Member of Academia Europaea and the Royal Society. He is also a fellow of the American Academy of Arts and Sciences, the Royal Academy of Engineering (U.K.), and the Royal Society of Edinburgh. He received a Guggenheim Fellowship in 2002 and the IEEE Education Medal in 2005. Recent recognition of his work includes the 2014 URSI Booker Gold Medal, and honorary doctorates from several universities in Asia and Europe. In 1990, he served as the President of the IEEE Information Theory Society, and from 2004 to 2007, as the Editor-in-Chief of the \textsc{IEEE Transactions on Information Theory}.
\end{IEEEbiography}

\end{document}